%% file: AssessmentOfDataDrivenModalDecompositionTechniques_V4.tex
\documentclass[review]{elsarticle}
\makeatletter
\def\ps@pprintTitle{%
  \let\@oddhead\@empty
  \let\@evenhead\@empty
  \let\@oddfoot\@empty
  \let\@evenfoot\@oddfoot
}
\makeatother

\usepackage{lineno,hyperref}
\modulolinenumbers[5]

\journal{Journal of \LaTeX\ Templates}

\graphicspath {{figures/}}
\input{customizations}
\usepackage{url}
\usepackage[utf8]{inputenc}
\usepackage{tabularx}
\usepackage{amsmath}
\usepackage{adjustbox}
\usepackage{float}

\begin{document}

\begin{frontmatter}

\title{Data--driven modal decomposition methods as\\ feature detection techniques for flow problems:\\ a critical assessment}

\newcommand{\orcidBB}{0000-0002-9993-0915} 
\newcommand{\orcidNG}{0000-0002-8099-0627} 
\newcommand{\orcidJGM}{0000-0002-7422-5320} 
\newcommand{\orcidSLCM}{0000-0003-3605-7351} 
\newcommand{\orcidEV}{0000-0002-1627-6883} 

\author[addressUPM]{B. Begiashvili \orcidlink{\orcidBB}}\corref{corresponding1}
\ead{beka.begiashvili@alumnos.upm.es}
\author[addressUPM,addressUPM2]{N. Groun \orcidlink{\orcidNG}}\corref{corresponding1}
\ead{gr.nourelhouda@alumnos.upm.es}
\author[addressUPM,addressCSC]{J. Garicano-Mena \orcidlink{\orcidJGM}}\corref{corresponding1}
\ead{jesus.garicano.mena@upm.es}
\author[addressUPM,addressCSC]{S. Le Clainche \orcidlink{\orcidSLCM}}\corref{corresponding1}
\ead{soledad.leclainche@upm.es}
\author[addressUPM,addressCSC]{E. Valero\orcidlink{\orcidEV}}
\ead{eusebio.valero@upm.es}
%
%
\address[addressUPM]{ETSI Aeron\'autica y del Espacio -- Universidad Polit\'ecnica de Madrid, 28040 Madrid, Spain}
\address[addressUPM2]{ETSI Telecomunicación  -- Universidad Politécnica de Madrid, 28040 Madrid, Spain}
\address[addressCSC]{Center for Computational Simulation (CCS), 28660 Boadilla del Monte, Spain}

\begin{abstract}
Modal decomposition techniques are showing a fast growth in popularity for their good properties as data-driven tools. There are several modal decomposition techniques, yet Proper Orthogonal Decomposition  (\POD) and Dynamic Mode Decomposition (\DMD) are considered the most demanded methods, especially in the field of fluid dynamics. Following their magnificent performance on various applications in several fields, numerous extensions of these techniques have been developed. In this work we present an ambitious review comparing eight different modal decomposition techniques, including most established methods: \POD, \DMD~and Fast Fourier Trasform (\FFT), extensions of these classical methods: based on time embedding systems, Spectral POD (\SPOD) and Higher Order DMD (\HODMD), based on scales separation, multi-scale POD (\mPOD), multi-resolution DMD (\mrDMD), and based on the properties of the resolvent operator, the data-driven Resolvent Analysis (\RA). The performance of all these techniques will be evaluated on three different testcases: the laminar wake around cylinder, a turbulent jet flow, and the three dimensional wake around cylinder in transient regime. First, we show a comparison between the performance of the eight modal decomposition techniques when the datasets are shortened. Next,  all the results obtained will be explained in details, showing  both the conveniences and inconveniences of all the methods under investigation depending on the type of application and the final goal (reconstruction or identification of the flow physics).  In this contribution we aim on giving a -as fair as possible- comparison of all the techniques investigated. To the authors knowledge, this is the first time a review paper gathering all this techniques have been produced, clarifying to the community what is the best technique to use for each application.        

\end{abstract}


\end{frontmatter}

%

\section{Introduction}\label{sec_introduction}
Nowadays, modal decomposition techniques have shown a fast growth in popularity in several fields and applications (aerospace and wind engineering, oil and gas industry, medicine...). These types of techniques have become very popular in the field of fluid dynamics, as they grant the  understanding of complex flows as well as allowing the reconstruction of diverse, useful information related to the flow physics.\\
In fluid mechanics there are two relevant modal decomposition techniques due to their tested good performance in a wide range of applications: Proper Orthogonal Decomposition (\POD) and Dynamic Mode Decomposition (\DMD). Several authors put a great effort in increasing the robustness of these methods with the aim also at widening their range of applications, hence they develop new variants and extensions of \DMD~and \POD~methods. To name a few: (i) snapshot \POD~\cite{sirovich1987turbulence,sirovich1987low}, this algorithm, which is based on the use of snapshots,  can be said to be an exact symmetry of classic \POD~\cite{tropea2007springer}, as it  involves correlations in time and averaging in space, in contrast to the classic \POD, which has correlations in space and averaging in time. When compared to classical \POD, which has limitations because of computational intensiveness, snapshot \POD~is sufficiently fast and computationally efficient. (ii) Extended \POD~(EPOD)~\cite{boree2003extended}, which is a generalization of the standard \POD, where it uses the temporal basis of one variable to find the modes of any data ensemble synchronized with the data ensemble used for the decomposition. This method have been mainly used, and found efficient in extracting information on correlated events  and (iii) cross \POD~(CPOD)\cite{cavalieri2021cross}, a quite recent extension of the classical \POD. The technique, which was developed in order to optimally decompose the trace of cross-covariances of flow fluctuations,  decomposes the flow in terms of modes that are ranked according to the cross correlation they carry. This technique shows a promising direction in the application to generalized Reynolds stresses. Regarding improvements of the classical \DMD, we mention the following: (i) extended \DMD~(EDMD)\cite{williams2015data}, which ameliorates the standard \DMD~by including a dictionary of observables, which spans a finite dimensional subspace on which we can approximate the Koopman operator and this will allow the method to capture more complex behavior. (ii) Sparsity promoting \DMD~(spDMD)\cite{jovanovic2014sparsity}, which uses sparse estimation and convex optimization techniques in order to end up with fewer, but more important modes and (iii) \DMD~with control (DMDc)\cite{proctor2016dynamic}, where this technique employs both measurements of the system and an orthogonal complement of control inputs to extract the underlying dynamics. \\  
In this contribution we are presenting a review comparing eight modal decomposition techniques, which can be categorized into three groups: (i) classical methods including Proper Orthogonal Decomposition (\POD)~\cite{sirovich1987turbulence}, Dynamic Mode Decomposition (\DMD)~\cite{schmid_dmd2008} and Fast Fourier Transform (\FFT)~\cite{cooleyTukeyFFT}, (ii) improved methods based on time-embedding systems:  Spectral Proper Orthogonal Decomposition  (\SPOD)~\cite{towneEtAlSPOD} and Higher Order Dynamic Mode Decomposition (\HODMD)~\cite{leClaincheHODMDSIAM2017} and (iii) multi-resolution methods capable to differentiate fast and slow scales: multi-scale POD (\mPOD)~\cite{mendezEtAlJFM2019} and multi-resolutional \DMD~(\mrDMD)~\cite{kutz2016multiresolution}. Finally we close with the latest modal decomposition technique, the Resolvent Analysis method (\RA)~\cite{jovanovic2005componentwise}, with the aim at showing the similarities between the resolvent modes and the DMD and POD modes. To the authors' knowledge, this is the first time all these techniques have been investigated and compared in one review paper. However, our objective is not to describe all the methods in details, but to make explicit the undefined framework underlying those methods.\\
These methods have been selected due to their strong potential for patterns identification in a wide range of applications. For instance, the \POD~for example was one of the first techniques to be employed in the field of fluid mechanics, where it was introduced in the context of turbulence by Lumley in 1967~\cite{lumley1967structure,lumley1981coherent} and continued to cover numerous application till this day. Smith \textit{et al.}~\cite{smith2005low} used \POD~to  build low-dimensional models for turbulent fluid flows. Willcox \textit{et al.}~\cite{willcox2002balanced} combined \POD~with concepts from balanced realization theory \cite{moore1981principal},  for the reduction
of  high-order systems, including a CFD model that describes the unsteady linearized motion of a two-dimensional airfoil. Zhang \textit{et al.}~\cite{zhang2021unsteady} leveraged \POD~to study the internal unsteady flow structure in a centrifugal pump for its advantages on turbulence analysis.\\
Another well-known technique is the \DMD. Right after introducing the \DMD~algorithm in 2010, Schmid \textit{et al.}~\cite{schmid2011applications} published a new work in order to  demonstrate the potential of the \DMD~technique, where they showed the robustness of their technique on two examples: (i) the decomposition of Schlieren snapshots of a helium jet and (ii) the decomposition of time-resolved PIV-measurements of an unforced and forced jet. Rowely \textit{et al.}~\cite{rowley2009spectral} aimed to describing the global behavior of complex nonlinear flows using the spectral analysis of
the Koopman operator, where they recruit the \DMD~to find the Koopman modes. The results demonstrated the capacity of the technique in capturing the dominant frequencies and elucidating the associated spatial structures. Kutz \textit{et al.}~\cite{kutz2016dynamic} produced a detailed, information rich book on the \DMD~algorithm. The book developed the fundamental theoretical foundations of \DMD~and highlighted many of its  applications. A few years earlier, Garicano-Mena \textit{et al.}~\cite{garicano2019composite} have also used the \DMD~algorithm to investigate both standard and actuated turbulent channel databases generated by  Direct Numerical Simulation (DNS), in order to explore the existence of flow features linked to drag reduction and the possibility of learning how those structures could be modified to better understand the efficiency of drag reduction strategies if they did exist. With a different approach, Barros \textit{et al.}~\cite{barros2022dynamic} worked on enabling \DMD~to extract features from observations with different mesh topologies and dimensions, such as those found in  adaptive mesh refinement/coarsening (AMR/C) simulations, as well as evaluating the \DMD~efficiency to reconstruct the dynamics and some relevant quantities of interest.\\
Similarly, extensions of the previously mentioned techniques are quite popular as well. A recent extension of \POD~is Spectral \POD~(\SPOD), which was developed in order to overcome the difficulties classical \POD~faces when the relevant coherent structures occur at low energies or at multiple frequencies, which is often the case. This algorithm, which can be applied to both spatially and temporally resolved data, has already covered several applications in fluid dynamics. Shahram \textit{et al.}~\cite{karami2018analysis} employed \SPOD~to study the spatio-temporal dynamics of the coherent structures in an under-expanded supersonic
impinging jet, where the \SPOD~decomposes the flow filed allowing the illustration of the forward and backward paths of the feedback loop, representing two things: (i) the hydrodynamic instability in the shear layer of the jet and (ii) the acoustic wave propagating in the medium. Leandra \textit{et al.}~\cite{abreu2020spectral} also used \SPOD~to identify energetically dominant coherent structures in turbulent pipe flows, using direct numerical simulations, performed with a high-order spectral-element method. Meanwhile Akhil \textit{et al.}~\cite{nekkanti2021frequency} used a large-eddy simulation data of a turbulent jet to demonstrate the applicability of the \SPOD~algorithm in four different application: low-rank reconstruction, denoising, frequency–time analysis and prewhitening. All the approached strategies of this work gave satisfactory results.\\ 
Higher order dynamic mode decomposition (\HODMD), is also a recent extension of one of the most known modal decomposition techniques, the \DMD. Even though it has been only five years since this algorithm was developed, it has covered several applications in the fluid dynamics field~\cite{ZHOU2021104545} and even reached biology ~\cite{le2019prediction} and the medical field~\cite{GROUN2022105384,GROUN2022106317}. Le Clainche \textit{et al.}~\cite{leClaincheHODMDSIAM2017} have demonstrated the efficiency of the \HODMD~algorithm on several applications including studying the flow structures of a zero-net-mass-flux (ZNMF) jet \cite{LeClaincheVegaSoria17}, analyzing turbulent flow of an elastoviscoplastic fluid \cite{leClaincheHODMDChannel} and to identify cross-flow instabilities \cite{LeClaincheHanFerrer19}. Similarly, M\'endez \textit{et al.} \cite{Mendezetal2021} introduced a new data processing method, based on \HODMD, to analyze actual flight test experimental data. Their proposed technique provides useful and interesting information for predicting flutter and proved to be suitable for the analysis of flight test data in real time.\\

Multi-scale \POD~(\mPOD)~\cite{mendezEtAlJFM2019} and multi-resolution \DMD~(\mrDMD)~\cite{kutz2016multiresolution} have also become very popular during the last days. Tang \textit{et al.}~\cite{tang2020tomographic} investigated the response of the large- and small-scale structures on the dynamic cylindrical element using \mPOD, where after employing the \mPOD~for data processing, the modes obtained by the technique were used to construct the large- and small-scale structures. Procacci \textit{et al.}~\cite{procacci2022multi} examined the flow field dynamics of bluff-body stabilized swirling and non-swirling flames using the \mPOD~algorithm. By applying this method, they were able to identify the main flow patterns in the velocity field and isolate coherent structures linked to various flow instabilities. Esposito \textit{et al.}~\cite{espositomulti} used the \mPOD~technique for the analysis of cavitation instabilities in water and liquid nitrogen. In this work, the authors highlighted the advantage of the \mPOD~in providing band-limited modes preserving anyway a good convergence unlike standard \POD. Climaco \textit{et al.}~\cite{climaco2021multi} proposed an approach for damage detection in wind turbine gearboxes using the \mrDMD~technique. Gearbox vibration signals generated under varying load conditions were analyzed using \mrDMD, which resulted the identification of important features related to the damage. Gonzales \textit{et al.}~\cite{gonzales2022multi} created a continuous-time model of the pressure profile over the fluttering airfoil using \mrDMD in order to provide a higher temporal resolution information about the system. The \mrDMD~algorithm was able to isolate the behavior of an oscillating shock wave over the surface of the fluttering airfoil, as well as  determine the frequency of the amplitude oscillations of the shock wave. \\
The last technique we discuss is the Resolvent Analysis (\RA). Before developing the data-driven \RA~by Herrmann \textit{et al.} in 2021 \cite{dmdResolventJFM}, the \RA~was already making an impact in the field of fluid mechanics, trying to conduct reserach towards new applications of flow control~\cite{jovanovic2005componentwise,luhar2014opposition}. Luhar \textit{et al.}~\cite{luhar2014structure} reformulated the resolvent analysis to generate predictions for the fluctuating pressure field in turbulent pipe flow. They used their proposed approach to show how the obtained response modes reconcile many of the key relationships among the velocity field, coherent structures and high-amplitude wall-pressure events. Yeh \textit{et al.}~\cite{yeh2019resolvent} employed the \RA~to design active control techniques for separated flows
over an airfoil. They also considered the use of a temporal filter to limit the time horizon and conducted a global resolvent analysis on the baseline turbulent mean flows to identify the actuation frequency and wavenumber that provide large perturbation energy amplification (see also \cite{sharma2009perturbation,chavarin2020resolvent}).
\\

The rest of the work is organized as follows: Section~\ref{sec_methodology} summarizes algorithms for these different techniques. Section~\ref{sec:NumSim} introduces and describes the test cases we are evaluating the algorithms on. A comparison to test the convergence of the method as function of the number of snapshots is presented in Section~\ref{sec_methodComparison}, followed by the main, detailed results, which are presented in Section~\ref{sec:Results}. Finally, Section~\ref{sec_conclusions} presents main conclusions of the work.\\

\section{Methodology}\label{sec_methodology}

In this section we present the different data analysis methods considered in this work. 
Those methods can be immediately separated into two major, distinct categories:
\POD-related~methods and \DMD-related methods. 
Regarding the upcoming discussion, however, one should bear in mind that strong connections exist between the methods inside a given category.
Moreover, links can also be established between both families of methods.
These relationships are perhaps better revealed if one resorts to matrix factorization based interpretations of the algorithms, see \eg~in Refs.~\cite{schmid_dmd2008,kutzEtAlBook}.

Consider a sequence of $n_p$-dimensional instantaneous flow fields $\mathbf{v}_j$,  (\eg, system states, snapshots or frames) indexed from $1$ to $n_t$ and that have been acquired at the uniform sampling rate $\Delta t^s=t_{j+1}-t_j$. 
Note that $n_p$ accounts for the number of spatial locations $n_s$ (grid points, pixels) and the number of flow variables considered $n_{vars}$ (velocity components, pressure, \ldots), so that $n_p = n_s\times n_{vars}$; by \textit{reshaping} the snapshots as column vectors $\mathbf{v}_j\in\mathbb{R}^{n_p}$, a \textit{data matrix} is constructed:
\begin{eqnarray}\label{eq_dataMatrixDefintion}
\mathbf{V}_1^{n_t} = \left[\mathbf{v}_1,\, \mathbf{v}_2,\,\ldots,\, \mathbf{v}_{n_t-1},\, \mathbf{v}_{n_t}\right]\in\mathbb{R}^{n_p\times n_t}.
\end{eqnarray}
\noindent Most fluid dynamics applications, be these numerical or experimental, 
lead to data matrices with $n_t\ll n_p$, termed Tall \& Skinny (\TS) matrices.
Notation-wise, we will often omit the sub and superindexes for the data matrix whenever they are clear from the context.

It is not uncommon to consider the associated dataset obtained by subtracting from each of the temporal snapshots the temporally averaged field, this is, the mean flow, \ie~$\bar{\mathbf{v}}=\frac{1}{n_t}\sum\limits_{j=1}^{n_t}\mathbf{v}_j$. 
Context often determines whether the original or the average-subtracted dataset are being considered, and hence we maintain the notation $\mathbf{V}_1^{n_t}$ for both.

Before dwelling into the specifics of the different methods considered, we recall that most data--driven modal analysis techniques can be interpreted as
matrix factorization strategies (see~\cite{schmid_dmd2008,kutzEtAlBook,bruntonKutzDataDrivenBook}, but also~\cite{mendezEtAlJFM2019}),
as in:
\begin{eqnarray}\label{eq_generalFactorization}
\mathbf{V}_1^{n_t} = \mathbf{A} \, \mathbf{B} \, \mathbf{C}^H = \sum\limits_{j=1}^{r} b_j\,\mathbf{a}_j\cdot \mathbf{c}_j^H,
\end{eqnarray}
\noindent where $\mathbf{A}\in\mathbb{C}^{n_p\times r}$
\footnote{Most \POD-related methods factorize the data matrix into real factors, 
whereas \DMD-related methods yield complex factors.
Those complex factors form nevertheless balanced complex-conjugated variable pairs, so that the real data matrix $\mathbf{V}_1^{n_t}$ is recovered.}, 
$\mathbf{B}\in\mathbb{C}^{r\times r}$, 
$\mathbf{C}\in\mathbb{C}^{n_t\times r}$ and $H$ is the complex-conjugate operator.
A diagonal structure is usually enforced for $\mathbf{B}$, 
and the dimension $r$ is chosen so that $1\leq r\leq \min(n_p,\,n_t)$.
Since $\mathbf{a}_{j}\in\mathbb{C}^{n_p}$,
matrix $\mathbf{A}$ is related to the \textit{spatial dimension} of the dataset
and thus $\mathbf{A}$ is sometimes referred as the \textit{topos} matrix.
In the same manner, $\mathbf{c}_{j}\in\mathbb{C}^{n_t}$,  and thus related to the \textit{temporal dimension};
matrix $\mathbf{C}$ is called accordingly the \textit{chronos} matrix.

In what follows, we review first the basic \POD, \DMD~ and \FFT~techniques, 
stressing out the connections between them.
Next, we address several extensions of those methods. 
The first extension leverages data redundancy either 
in the spectral domain, by averaging spectra from shorter subsequences (in the spirit of the Welch method~\cite{welchIEEE1967}), 
or in the temporal domain through time--lagging of the data snapshots;
in both cases, improved analysis  techniques are obtained.
Spectral Proper Orthogonal Decomposition~\cite{towneEtAlSPOD} and Higher Order Dynamic Mode Decomposition~\cite{leClaincheHODMDSIAM2017} are representatives of this approach.
Next, a multi--scale approach for \POD,
and a multi--resolution interpretation of \DMD~are introduced.
Finally, resorting to an input--output perspective in the space of states~\cite{jovanovicAnnualReview} 
allows to introduce the data--driven, \DMD--based Resolvent Analysis~(\RA) technique described in~\cite{dmdResolventJFM}.

Table~\ref{tab_summary} lists the different methods analyzed in this contribution.
For each, one or two fundamental references are provided.
The table also credits the origin of the implementations for the different methods, at least when it is available.

%

\begin{table}[h]
\centering
\caption{\label{tab_summary} Summary of the data--analysis methods considered in this work, principal references and availability.}
\begin{adjustbox}{max width=\textwidth}
		\begin{tabular}{c c c c}
		\toprule
			Method      & Major  references     &     & Available from \\
			\midrule                                                      
			\POD        & \cite{volkweinPODLectureNotes,Berkooz_POD}              &     & \texttt{Matlab} commands \texttt{svd}, \texttt{eig}      \\
			\DMD        & \cite{schmid_dmd2008,jovanovic_SPDMD}                   &     & \url{http://www.ece.umn.edu/users/mihailo/software/dmdsp/}      \\
			\FFT        & \cite{cooleyTukeyFFT,bruntonKutzDataDrivenBook}         &     & \texttt{Matlab} native commands \texttt{fft}, \texttt{fft2}  \\
			\SPOD       & \cite{towneEtAlSPOD,SPODGuideAIAAJ}                     &     & \url{https://nl.mathworks.com/matlabcentral/fileexchange/65683-spectral-proper-orthogonal-decomposition-spod}\\
			\HODMD      & \cite{leClaincheHODMDSIAM2017,bookLeClaincheHODMD}      &     & \url{https://short.upm.es/q8rip} \\
			\mPOD       & \cite{mendezEtAlJFM2019, 2019MendezEtAlExpThermalFluid} &     & \url{https://github.com/mendezVKI/MODULO}  \\
			\mrDMD      & \cite{kutz2016multiresolution, kutzEtAlBook}                         &     & \url{https://github.com/kdmarrett/dmd}  \\
			\RA         & \cite{dmdResolventJFM}                                  &     & \texttt{N.A.}/Own implementation  \\
			\bottomrule
		\end{tabular}
\end{adjustbox}
\end{table}

\subsection{The Proper Orthogonal Decomposition methods}\label{subsec_basicPOD}
We begin this necessarily brief review of classical Proper Orthogonal Decomposition methods by acknowledging that
many related methods are referred to as \POD~in the bibliography.
Principal Component Analysis or Karhunen-Lo\`eve decomposition are also terms applied to these techniques.

Reference~\cite{volkweinPODLectureNotes} presents a good, integrative description of
three different yet interrelated \POD~algorithms. 
We cover here two of them, as they will be useful in the posterior discussion. 

A first strategy consists in 
the direct application of the (economy--sized) Singular Value Decomposition (\SVD) to 
the average substracted dataset:
\begin{eqnarray}\label{eq_spatialReduction0}
    \mathbf{V}_1^{n_t} \overset{SVD}{=} \mathbf{L}_0\;\mathbf{S}_0\;\mathbf{R}_0^T = 
    \sum\limits_{j=1}^{r} s_{0,j}\,\mathbf{l}_{0,j}\cdot \mathbf{r}_{0,j}^T. 
\end{eqnarray}


Matrix $\mathbf{S_0}$ contains as diagonal entries the non--negative and decreasing singular values $s_{0,j}$,
whereas the real matrices $\mathbf{L}_0$ and $\mathbf{R}_0$, which are orthogonal,  
have as their columns $\mathbf{l}_{0,j}$ and $\mathbf{r}_{0,j}$ the left and right singular vectors. Note that superindex $T$ stands for the transposition operator.
Following Refs.~\cite{rowleyDawson_annualReview_ModelReduction,schmid_dmd2008,jovanovic_SPDMD,kouLeClaincheHODMDCriterion}, 
the left singular vectors $\mathbf{l}_{0,j}$ are identified with the \POD~modes.

The second strategy, which for historical reasons is termed the \textit{method of snapshots}~\cite{sirovich1987turbulence},
considers the temporal correlation matrix $\mathbf{K}\equiv \left(\mathbf{V}_1^{n_t}\right)^T\;\mathbf{V}_1^{n_t}$, and its eigenvalue decomposition:
\begin{eqnarray}\label{eq_pod_strategy2}
\mathbf{K}\;\mathbf{R}_0 = \mathbf{R}_0\;\mathbf{D}_{s^2}, 
\end{eqnarray}
\noindent where the singular values are retrieved as 
the square root of the diagonal entries of $\mathbf{D}_{s^2}$, 
\ie, $s_{0,j} = \sqrt{D_{s^2, j,j}}$, and 
the \POD~modes are given by:
\begin{eqnarray}\label{eq_pod_strategy2_modes}
\mathbf{L}_0=\mathbf{V}_1^{n_t}\,\mathbf{R}_0\,\mathbf{D}_{s^2}^{-1/2}.
\end{eqnarray}

Observe how the data matrix factorization given by Eq.~\eqref{eq_spatialReduction0} allows 
to somehow \textit{separate} the spatial components from the temporal components of the database. 
Accordingly, one can define the \textit{scaled chronos matrix} as:
\begin{eqnarray}\label{eq_scaledChronosMatrix}
\mathcal{C}_1^{n_t}\equiv \mathbf{S}_0 \, \mathbf{R}_0^T=\sum\limits_{j=1}^{r}\,s_{0,j}\,\mathbf{e}_j\cdot\mathbf{r}_{0,j}^T, 
\end{eqnarray}
where $\mathbf{e}_j$ is the $j$-th unit vector in $\mathbb{R}^{r}$. 
This matrix will be useful when presenting the Higher Order \DMD~method in \S~\ref{subsec_HODMD}.

Finally, we point out that the non--increasing singular values in $\mathbf{S}_0$
have a relationship with the Frobenius norm of $\mathbf{V}_1^{n_t}$, 
namely $\|\mathbf{V}_1^{n_t}\|_F=\sqrt{\sum_j s_{0,j}^2}$.
Note also that the upper summation limit in Eq.~\eqref{eq_spatialReduction0} needs to be $r=\min{(n_p, n_t)}$.
The properties of the \SVD~decomposition~\cite{volkweinPODLectureNotes},
guarantee that choosing $1\leq r'\leq r$ leads to the $r'$-rank optimal representation of the dataset.
The $r'$ parameter,  often referred to as  \textit{spatial complexity}~\cite{LeClaincheVegaJNLS18}, can be chosen either  directly (simply by  specifying  an integer value), 
or indirectly, \ie,  adjusting $r'$ in order to meet a given tolerance $\varepsilon$, as in:

\begin{eqnarray}\label{eq_truncationStrategy}
\frac{\|\mathbf{V}_1^{n_t} - \sum\limits_{j=1}^{r'}s_{0,j}\,\mathbf{l}_{0,j}\cdot \mathbf{r}_{0,j}^T\|_F}
{\sqrt{\sum\limits_{j=1}^{r}s_{0,j}^2} } = \sqrt{\frac{\sum\limits_{j=r'}^{r}s_{0,j}^2}{\sum\limits_{j=1}^{r}s_{0,j}^2}} \leq  \varepsilon.
\end{eqnarray}

\subsection{The Dynamic Mode Decomposition method}\label{subsec_basicDMD}
The first implementations of the, now established, Dynamic Mode Decomposition methods 
can be traced back to Ref.~\cite{rowleyEtAlJFM2009} (the \textit{companion--matrix} \DMD~of Rowley \etal) and
to Refs.~\cite{schmid_dmd2008,schmid2008dynamic} (the \textit{similarity--transformation} \DMD~of  Schmid). 
In this work, and for brevity, we consider mainly implementations derived from the work of Ref.~\cite{schmid_dmd2008}.
The classical \DMD~method assumes a linear relationship between consecutive snapshots
\begin{eqnarray}\label{eq_koopmanDMD1}
\mathbf{v}_{k+1}=\mathcal{A}^{h}\,\mathbf{v}_{k},
\end{eqnarray}

\noindent where the linear operator $\mathcal{A}^h$ can be interpreted as 
a state transition matrix in discrete time~\cite{jovanovicAnnualReview}.
Eq.~\ref{eq_koopmanDMD1} is sometimes termed the \textit{Koopman assumption}~\cite{bookLeClaincheHODMD}.
Defining the  partial subsequences $\mathbf{X}\equiv\mathbf{V}_1^{n_t-1}$ and 
$\mathbf{Y}\equiv\mathbf{V}_2^{n_t}$, allows to write the former in compact form:
\begin{eqnarray}\label{eq_dmd_3}
\mathbf{Y}=\mathcal{A}^{h}\;\mathbf{X}.
\end{eqnarray}
Next, \SVD~of the first subsequence is performed, $\mathbf{X} \overset{SVD}{=} \mathbf{L}_0\;\mathbf{S}_0\;\mathbf{R}_0^T$, \cfr~Eq.~\eqref{eq_spatialReduction0}, 
and recall that, at this stage, one could consider a reduced representation 
of the input data subsequence $\mathbf{X}$ simply by retaining $r'\leq r=n_t-1$ singular values,
according to Eq.~\eqref{eq_truncationStrategy}.

Using the \SVD~of the $\mathbf{X}$ matrix into Eq.~\eqref{eq_dmd_3}
allows to build a reduced matrix $\bar{\mathbf{A}}$, defined as:
\begin{equation}\label{eq_dmd_4}
\bar{\mathbf{A}} \equiv \mathbf{L}_0^T\,\mathcal{A}^{h}\, \mathbf{L}_0=\mathbf{L}_0^T\,\mathbf{Y}\,\mathbf{R}_0\,\mathbf{S}_0^{-1}.
\end{equation}
  
\noindent The reduced matrix $\bar{\mathbf{A}}$ is the projection of 
the matrix $\mathcal{A}^{h}$ onto the space linearly generated by the columns of $\mathbf{L}_0$~\cite{schmid_dmd2008}.
The \DMD~method operates under the assumption that the projected matrix $\bar{\mathbf{A}}$ 
conveys most of the information codified into operator $\mathcal{A}^{h}$.

Once the reduced matrix $\bar{\mathbf{A}}$ has been calculated, 
its (right) eigenvalue decomposition: 
\begin{eqnarray}
\bar{\mathbf{A}}\;\boldsymbol{\Psi} = \boldsymbol{\Psi}\;\boldsymbol{\Lambda}_{\mu},
\end{eqnarray}
\noindent offers the reduced \DMD~modes $\boldsymbol{\psi}_i$ as the columns of $\boldsymbol{\Psi}$;
the corresponding eigenvalues $\mu_i$ (the diagonal entries of $\boldsymbol{\Lambda}_{\mu}$) indicate
the temporal growth rates ($\sigma_i=\Re(\mu_i)$) and angular pulsation ($\omega_i=\Im(\mu_i)$). \\

The projected eigenmodes (namely, the dynamic modes) of  matrix $\mathcal{A}^h$ are recovered as
$\boldsymbol{\Phi}=\mathbf{L}_0\,\boldsymbol{\Psi}$.

Note also that the growth rates and frequencies in the complex half-plane can 
be recovered from the eigenvalues as:
\begin{eqnarray}\label{eq_discreteToContinuous}
\lambda_i=\log(\mu_i)/\Delta t^s.
\end{eqnarray}

Finally, note that the \DMD~decomposition allows to reconstruct the original data sequence as:
\begin{eqnarray}\label{eq_expansionDMD}
  \mathbf{v}(t_j) = \sum_{i=1}^{r}\alpha_i \, \boldsymbol{\phi}_i \mu_i^{ j\, \Delta\,t^s},
\end{eqnarray}
\noindent expression that can be recast in matrix form as:
\begin{eqnarray}\label{eq_expansionDMD_matrixForm}
  \mathbf{X} =  \boldsymbol{\Phi} \, \mathbf{D}_\alpha \, \mathbf{V}_\mu,
\end{eqnarray}
\noindent where $\mathbf{V}_\mu$ is a Vandermonde matrix whose columns are generated by 
the successive powers of the column vector $\left[\mu_1^j,\ldots, \mu_{r}^j\right]^T$, with $j=0,\ldots,n_t-1$; 
and $\mathbf{D}_\alpha$ is a diagonal matrix whose non-zero entries are to be determined. 
There are several strategies available to identify $\mathbf{D}_\alpha$;
we defer the discussion and comparison to \S~\ref{app_amplitudes}.
\noindent Regarding Eq.~\eqref{eq_expansionDMD}, note that if an upper limit $r''\le r'\le r$ is considered, 
the original data $\mathbf{X}$ is not  reconstructed, but approximated. 
The parameter $r''$ is termed the \textit{spectral complexity}~\cite{LeClaincheVegaJNLS18},
and can also be tuned directly or indirectly through an $\varepsilon$-condition similar to that in Eq.~\eqref{eq_truncationStrategy}.
\\

We review now an alternative definition of the \DMD~method 
that eases the connection with the Higher Order \DMD~method discussed in \S~\ref{subsec_HODMD}.
This  formulation, that follows~\cite{leClaincheHODMDSIAM2017}, 
establishes the Koopman assumption on subsequences built from the scaled chronos matrix,
namely:
\begin{eqnarray}\label{eq_DMDOnScaledChronosMatrix}
\mathcal{C}_2^{n_t}=\mathcal{A}_1^h\,\mathcal{C}_1^{n_t-1}.
\end{eqnarray}
The usual procedure is then applied, \ie, the first subsequence is decomposed as 
$\mathcal{C}_1^{n_t-1}\overset{SVD}{=} \mathbf{L}_1\,\mathbf{S}_1\,\mathbf{R}_1^T$,
and so on. The only difference is that the dynamic modes are now recovered as 
$\boldsymbol{\Phi}=\mathbf{L}_0\,\mathbf{L}_1\,\boldsymbol{\Psi}$. 
This alternative formulation allows to retrieve the \DMD~analysis 
at a reduced computational cost, provided the \SVD~decomposition is already available,~\cite{liEtAl2022ThetaDMD,GROUN2022105384}.

At this point, we can make a first comparison between the \POD~and the \DMD~factorizations,
according to the factorization model $\mathbf{V}_1^{n_t}=\mathbf{A}\,\mathbf{B}\,\mathbf{C}$ of Eq.~\ref{eq_generalFactorization}.
The methods differ in the nature of the respective topoi matrices:
the \POD~modes $\mathbf{L}_0$ are orthogonal to each other, 
whereas the \DMD~modes $\boldsymbol{\Phi}$ are not necessarily so.
The \POD~chronos matrix $\mathbf{R}_0$ contains 
all the range of frequencies present in the problem analyzed (\ie, frequencies are \textit{mixed}),
whereas each of the rows of the \DMD~$\mathbf{V}_\mu$ matrix represents 
a sinusoidal temporal variation at a distinct frequency $\omega_j$:
these rows are therefore orthogonal \textit{in time}.
Finally, we insist on the fact that 
the singular values $s_{0,j}$ are non--increasing, 
and have a direct relationship with the Frobenius norm of $\mathbf{V}_1^{n_t}$, 
whereas the amplitudes $\alpha_j$ do not. 
In~\ref{app_amplitudes} we comment on the different strategies of computing and selecting the amplitudes.

\subsection{Fourier analysis}\label{subsec_FFT}
As discussed in, \eg~Refs. \cite{bookStrangAppliedMath, bruntonKutzDataDrivenBook},
the Discrete Fourier Transform~(\DFT) of 
the average-subtracted dataset $\mathbf{V}_1^{n_t}$ is formally computed as the matrix--matrix product:
\begin{eqnarray}\label{eq_FFT}
\hat{\mathbf{V}} =\mathbf{V}_1^{n_t}\,\mathcal{F},
\end{eqnarray}
\noindent where matrix $\mathcal{F}\in\mathbb{C}^{n_t\times n_t}$ is 
a hermitian matrix ($\mathcal{F}^{H}=\mathcal{F}$) generated from the $n_t$ roots of unity.
Matrix $\mathcal{F}$ has a special structure that can be exploited to
efficiently compute the transformation using a Fast Fourier Transform method~\cite{cooleyTukeyFFT,bruntonKutzDataDrivenBook}.
Note also that, if a \SVD~of the dataset is available, the \FFT~can be applied to the scaled chronos matrix.

In this work, we consider the Power Spectral Density~(\PSD) of time-resolved datasets, computed as:
\begin{eqnarray}\label{eq_psd}
PSD = \hat{\mathbf{V}} \odot \hat{\mathbf{V}}^H,
\end{eqnarray}
\noindent where $\odot$ represents the Hadamard (elementwise) product. 

The Fourier transformation of the dataset also opens the way to conduct manipulations in the spectral space,
which is at the root of both Spectral \POD~and multi-scale \POD~methods:
these connections will be stressed in the corresponding sections. 
Regarding \DMD, Rowley \etal~established that conducting \DMD~on the average-subtracted dataset is equivalent to \DFT~\cite{chenTuRowleyDMDVariants}.

\subsection{Improving accuracy through redundancy: window-shift methods}
Analyzing the power spectral density of a infinitely long, periodic signal using \DFT~techniques
becomes easier as longer and longer temporal sequences are available.
In practice, however, seldom can signals be sampled for long enough.
Moreover, spectral estimates do not converge as the number of samples $n_t$ is increased:
the uncertainty of the estimate at each frequency is as large as the magnitude of the estimate itself.

One strategy to obtain converged estimates averages spectra over several realizations of the signal,
using \eg~the Welch method~\cite{welchIEEE1967}.
This strategy is at the root of the Spectral Proper Orthogonal Decomposition (\SPOD) method~\cite{towneEtAlSPOD}.
An equivalent effect can be obtained by resorting to the analysis of $d$ time-lagged snapshots, 
as done by the Higher Order Dynamic Mode Decomposition (\HODMD) method~\cite{leClaincheHODMDSIAM2017,bookLeClaincheHODMD}.
We review both methods next.

\subsubsection{The Spectral Proper Orthogonal Decomposition method}\label{subsec_SPOD}
Spectral Proper Orthogonal Decomposition is a spatio-temporal decomposition for 
statistically stationary data that relies on estimating first, 
and factorizing next the cross--spectral density (or \CSD) tensor at different frequencies.
In this work we limit ourselves to a succinct description of the \SPOD~method, 
mostly from the algorithmic point of view;
Refs.~\cite{towneEtAlSPOD,SPODGuideAIAAJ} provide 
a detailed account of the method and its application.
As a final comment, note that other data processing strategies also termed~\SPOD~
have been described in the literature, \eg~\cite{sieberEtAlSPOD,derMurEtAlSPOD}, but we do not cover them here.

\begin{figure}[h]
	\centering
	\includegraphics[height=7.0cm]{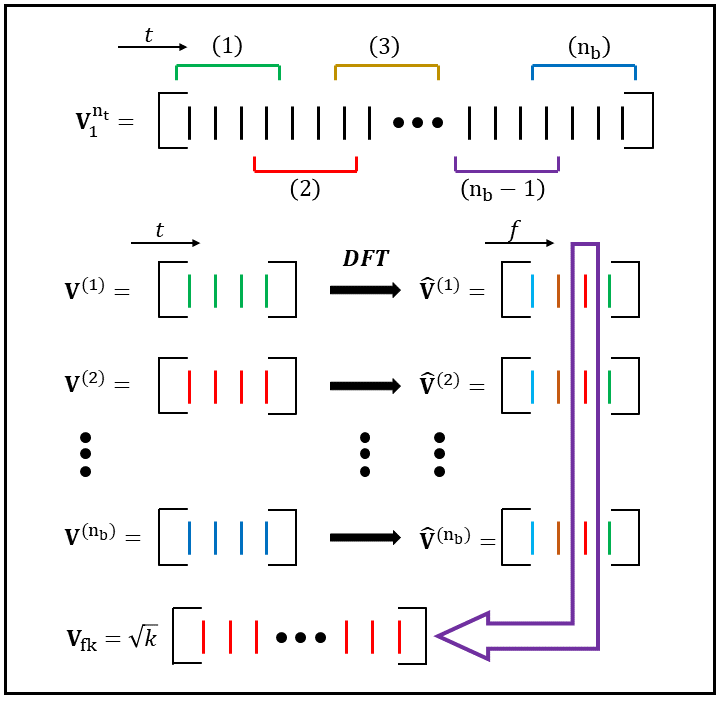}
	\caption{Schematic depiction of the method described in \cite{towneEtAlSPOD} for estimating \SPOD~modes.}\label{fig_spod_algorithm}
\end{figure}

Fig.~\ref{fig_spod_algorithm} is illustrative of the \SPOD~algorithm.
Estimates of the \CSD~tensor at a predefined set of frequencies $f_k$ can be obtained 
simply by subdividing the dataset into $n_b$ subsequences --or blocks-- 
of temporal length $n_t'$, $\mathbf{V}^b=\mathbf{V}_{k}^{k+n_t'}$, with $1\leq k\leq n_t-n_t'$ and $b=1,\ldots, n_b$. 
Each of these blocks, which might share (\textit{overlap}) with neighbouring blocks up to 
$n_o$ snapshots\footnote{Parameter $n_o$ is usually chosen so that a 50\% overlap between consecutive subsections is attained, \cite{SPODGuideAIAAJ}}, is Fourier transformed to $\mathbf{\hat{V}}^b$.

By observing that the $k$-th column for any of the transformed blocks $\mathbf{\hat{V}}^b$ 
is related to the same frequency $f_k$ (see Fig.~\ref{fig_spod_algorithm},
it is possible to estimate the \CSD~tensor at $f_k$ 
simply by grouping all those columns into a matrix $\mathbf{V}_{f_k}$ and operating:
\begin{equation}\label{eq_estimateOfCSD}
\mathbf{S}_{f_k}= \mathbf{V}_{f_k}^H\;\mathbf{W}\;\mathbf{V}_{f_k}.
\end{equation}
\noindent The formulation admits a weight matrix $\mathbf{W}$ that 
may serve to account for mesh stretching and/or window functions~\cite{volkweinPODLectureNotes,towneEtAlSPOD}.
In this work, we use the Hamming window function~\cite{SPODGuideAIAAJ}.
Note that the frequencies $f_k$ are defined a priori by the choice of $n_b$ and $n_o$, see \eg~Eq. 3.5 in \cite{towneEtAlSPOD}

The \SPOD~modes are obtained from the eigenvalue decomposition, \cfr~Eqs.~\eqref{eq_pod_strategy2} and~\eqref{eq_pod_strategy2_modes}:
\begin{eqnarray}
\mathbf{S}_{f_k}\,\boldsymbol{\Theta}_{f_k}= \boldsymbol{\Theta}_{f_k}\,\boldsymbol{\Lambda}_{f_k}, 
\mbox{ followed by }
\mathbf{L}_{0,f_k}=\mathbf{V}_{f_k}\,\boldsymbol{\Theta}_{f_k}\,\boldsymbol{\Lambda}_{f_k}^{-1/2}.
\end{eqnarray}
\noindent Thus, the \SPOD~method follows the algorithmic strategy of Eq.~\ref{eq_pod_strategy2}.
As such, the modal energies at the frequency $f_k$  is given by the diagonal entries of ${\Lambda}_{f_k}^{1/2}$. 
Those modal energies are non--increasing, and  thus enable \textit{sorting} the contribution of the different \SPOD~modes.

\subsubsection{The Higher Order Dynamic Mode Decomposition method}\label{subsec_HODMD}

The Higher Order \DMD~method (\HODMD) leverages also the notion of redundancy, but in time.
The interested reader can consult the extensive bibliography on the topic, \eg~Refs.~\cite{leClaincheHODMDSIAM2017,LeClaincheVegaSoria17} or the monograph Ref.~\cite{bookLeClaincheHODMD}.



\begin{figure}[h]
	\centering
	\includegraphics[height=7.0cm]{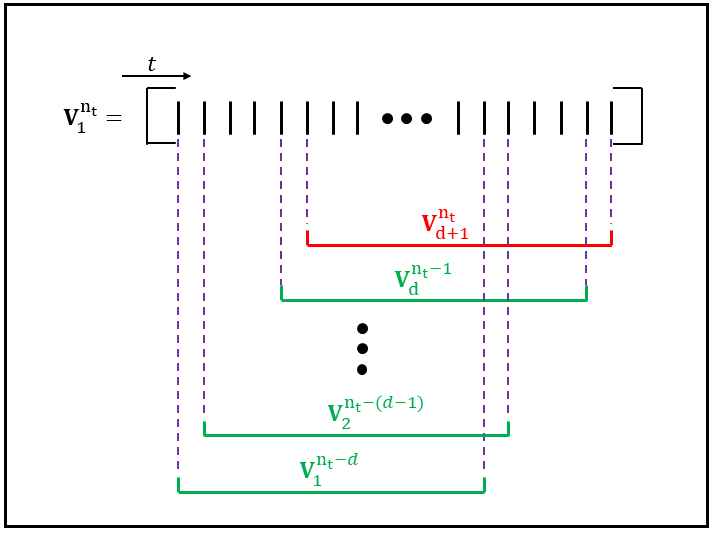}
	\caption{Schematic depiction of the \HODMD~method.}\label{fig_hodmd_algorithm}
\end{figure}

In order to introduce the \HODMD, the Koopman assumption of Eq.~\ref{eq_koopmanDMD1} is extended 
to include $d$ time--lagged snapshots\footnote{Subindex $1$ and superindex $h$ has been removed from the operator $\mathcal{A}^{h}_{1, j}$ to ligthen the notation.}:
\begin{eqnarray}\label{eq_koopmanDMDd}
\mathbf{v}_{k+1}=\sum\limits_{j=1}^{d}\mathcal{A}_j\,\mathbf{v}_{k-j}.
\end{eqnarray}
\noindent This relation is sometimes termed the \textit{higher order} Koopman assumption.
For reasons of computational efficiency, it is better to write Eq.~\ref{eq_koopmanDMDd} 
in terms of the columns of the scaled chronos matrix:
\begin{eqnarray}\label{eq_koopmanDMDdChronos}
\mathbf{c}_{k+1}=\sum\limits_{j=1}^{d}\mathcal{A}_{j}\,\mathbf{c}_{k-j}.
\end{eqnarray}
\noindent Note how setting $d=1$ in either Eq.~\ref{eq_koopmanDMDd} or~\eqref{eq_koopmanDMDdChronos} leads to the \textit{classical} \DMD~method. The matrix counterpart of Eq.~\ref{eq_koopmanDMDdChronos}, see Fig.~\ref{fig_hodmd_algorithm}:
\begin{equation}
\mathbf{C}_{d+1}^{n_t}=
\mathcal{A}_1 \mathbf{C}_1^{n_t-d} +
\mathcal{A}_2 \mathbf{C}_2^{n_t-d+1} + \ldots +
\mathcal{A}_d \mathbf{C}_d^{n_t-1},
\end{equation}

\noindent Mathematical manipulations lead to a relation based on a companion-like block matrix:
\begin{equation}
\begin{bmatrix}
\mathbf{C}_2^{n_t-d+1}\\
\mathbf{C}_3^{n_t-d+2}\\
\vdots\\
\mathbf{C}_d^{n_t-1}\\
\mathbf{C}_{d+1}^{n_t}
\end{bmatrix}
=
\begin{bmatrix}
0            &\mathbf{I}   &0            &\ldots&0              \\
0            &0            &\mathbf{I}   &\ldots&0              \\
\vdots       &\vdots       &\vdots       &\ddots&\vdots         \\
0            &0            &0            &\ldots&\mathbf{I}     \\
\mathcal{A}_1&\mathcal{A}_2&\mathcal{A}_3&\ldots&\mathcal{A}_{d}
\end{bmatrix}\;
\begin{bmatrix}
\mathbf{C}_1^{n_t-d}\\
\mathbf{C}_2^{n_t-d+1}\\
\vdots\\
\mathbf{C}_{d-1}^{n_t-2}\\
\mathbf{C}_{d}^{n_t-1}
\end{bmatrix},
\end{equation}
\noindent which can be written compactly as:
\begin{equation}\label{eq_modkoopmanmat}
\widetilde{\mathbf{C}}_2^{n_t-d+1}=
\widetilde{\mathbf{A}}\;
\widetilde{\mathbf{C}}_1^{n_t-d}.
\end{equation}
\noindent  Note how this last equation is formally similar to Eq.~\eqref{eq_DMDOnScaledChronosMatrix}.
Indeed, from here on the \HODMD~follows the classical \DMD~algorithm.
Accordingly,  matrix $\widetilde{\mathbf{A}}$  is never explicitly formed;
the algorithm proceeds instead by the \SVD:
\begin{equation}\label{eq_hodmdFirstSVD}
    \widetilde{\mathbf{C}}_1^{n_t-d}\overset{SVD}{=} \widetilde{\mathbf{L}}_1\,\widetilde{\mathbf{S}}_1\,\widetilde{\mathbf{R}}_1^T.
\end{equation}
This in turn serves to build the projected matrix (\cfr~Eq.~\eqref{eq_dmd_4}):
\begin{eqnarray}
\bar{\mathcal{A}}=\widetilde{\mathbf{C}}_2^{n_t-d+1}\,\widetilde{\mathbf{R}}_1\widetilde{\mathbf{S}}_1^{-1}\,\widetilde{\mathbf{L}}_1^T.
\end{eqnarray}
\noindent The eigenvalue decomposition of this last matrix, 
$\bar{\mathcal{A}}\,\mathbf{\Psi}  = \mathbf{\Psi}\,\boldsymbol{\Lambda}_{\mu}$, 
provides the \HODMD~modes $\mathbf{\Phi}=\mathbf{L}_0\,\widetilde{\mathbf{L}}_1\,\mathbf{\Psi}'$,
where $\mathbf{\Psi}'=\mathbf{\Psi}(1:r_1,:)$. 


We  remark that the \HODMD~method offers two complementary alternatives for the reduced representation of the dataset:
one when building the scaled chronos matrix in Eq.~\eqref{eq_scaledChronosMatrix} (associated to $r_1'$ and/or $\varepsilon_1$), and
the other with the \SVD~of matrix $\widetilde{\mathbf{C}}_1^{n_t-d}$ (through $r_2'$ and/or $\varepsilon_2$).

In this respect, for a sufficiently large number of snapshots $n_t$ (the common situation), 
whenever the spatial complexity $r'$ (the number of \POD/\SVD~modes retained after the dimensionality reduction carried out by the method, as function of the tolerance $\varepsilon_1$) is smaller than the spectral complexity $r''$ (the number of \DMD~modes retained in the \DMD~expansion Eq.~\eqref{eq_expansionDMD}, for $n_t > r'$), the high-order Koopman assumption completes the lack of spatial information (reduced from $n_t$ to $r'$). 
This explains the good performance of the \DMD~method in highly complex databases, when the noise (experiments) and/or the small flow scales (turbulence) are removed. 

\subsubsection{On the common traits of \SPOD~and~\HODMD~techniques\label{sec:RemarksSpodHodmd}}


From the previous discussion, it is evident that both \HODMD~and \SPOD~methods are improved versions of \DMD~and the method of snapshots \POD~techniques, respectively.
Despite obvious algorithmic differences between the \HODMD~and \SPOD~implementations presented so far,
one  should recognize  that both methods inherit their improved capabilities from the same underlying principle, 
which is the exploitation of data redundancy to better identify relevant features from the available data. 
\HODMD, on the one hand, begins with a data-dimensionality reduction step sustained on a first \SVD, and exploits the redundancy through a sliding window process. This process leads to an enlarged snapshot matrix with the data clean from noise, small flow scales, spatial redundancies/degeneracies or other undesirable artefacts; application of the standard \DMD~algorithm (including an additional dimensionality reduction step to alleviate the computer memory requirements)  allows then to identify the \DMD~modes, growth rates and frequencies. 
The \SPOD~method follows a different route to introduce redundancy on the frequency domain considering shorter, overlapping, Fourier-transformed window-sets that are also followed by a sliding window process. 
This step of the algorithm reminds the well-known Welch  method~\cite{welchIEEE1967},  
or, alternatively the \PSD~process, usually carried out when analysing experimental databases \cite{soria_leclanche}. 
This process leads to data subsets associated to specific frequency bins (say $f_k$) that are processed through the method of snapshots \POD, to yield finally a collection of \SPOD~modes associated to distinct frequencies, as in the \DMD~ method~\cite{towneEtAlSPOD}.

In cases with simple dynamics, \HODMD~and \SPOD~should provide similar results. However, in complex flows, encompassing a large number of spatio-temporal flow scales, both methods are complementary, generally retrieving modes of similar shape that contain relevant information about the main flow instabilities leading the flow. It is remarkable that \HODMD~provides the \DMD~modes, while \SPOD~provides a new set of modes, which could be considered a hybrid between \DMD~and \POD~modes: indeed, Ref.~\cite{towneEtAlSPOD} identifies the \SPOD~modes as ``optimally averaged \DMD~modes from an ensemble \DMD~problem for stationary flows''. In the analysis of complex flows, both methodologies require careful calibration. As it will be presented in Section \ref{sec:Results}, the number/size of the windows selected in both methodologies is crucial to properly identify the dynamics driving the flow, \ie, ~using too short windows will prevent both methods to identify low frequency relevant dynamics; alternatively, for complex datasets, a too small number of windows will negatively impact both methods' ability to filter out redundant frequencies. 

One of the main advantages of \HODMD~is that it is a method based on \DMD, which it is considered as a tool more efficient than \FFT: \DMD~is able to identify the main dynamics of the flow using smaller datasets than \FFT, which requires large databases to provide accurate results \cite{leClaincheHODMDSIAM2017}. Both, \FFT~and \DMD~follow Nyquist theorem, although in complex cases, it is known that \DMD~identifies the main frequencies of the flow in a database containing the 75\% of the period of the frequency \cite{chenTuRowleyDMDVariants}. Similarly to \FFT, \HODMD~is able to identify the main frequencies of the flow even in databases containing limited spatial information (\ie., experimental probes), but taking improvement from the advantages of \DMD~method. 
Also, \HODMD~selects the main frequencies of the flow automatically, based on the robustness of the results presented with the different calibration. Selecting specific frequencies to study the main flow dynamics enforces a prior knowledge of the flow studied, which could be a great inconvenient in the case of complex flows (\ie., turbulent channel flows \cite{leClaincheHODMDChannel,LeClaincheRostiBrandtJFM22}, flutter identification in flight test \cite{LeClaincheetalJAircraft18,Mendezetal2021}, identification of cross flow instabilities \cite{LeClaincheHanFerrer19}, predictions in lidar measurements~\cite{le2018wind} etc.
Finally, the application of the \HODMD~method is not constrained to statistically stationary data:
\HODMD~can be also applied to flow data acquired over transient regimes(\cite{LeClaincheetalFDR18,LeClaincheVegaPoF17}).
\subsection{Improving spectral estimates using multiresolution analysis}\label{subsec_MR}

\subsubsection{Multi-Scale Proper Orthogonal Decomposition}\label{subsec_MRPOD}



\noindent Multi-scale proper orthogonal decomposition (or \mPOD, introduced in Refs.~\cite{mendezEtAlExpTFS2018, mendezEtAlJFM2019}) combines 
multi-resolution analysis (\MRA) principles~\cite{bookStrangAppliedMath} with the method-of-snapshots \POD~technique (Eq.~\eqref{eq_pod_strategy2}). 
\MRA~allows to \textit{look} at the correlation matrix with different levels of detail.
This hierarchy of points of view allows to segregate the correlation matrix into contributions from different non-overlapping \textit{scales}, whereas the \POD~optimality properties allow to find the optimal basis for each scale.

\begin{figure}[H]
 	\centering
	\subfloat[]{\label{fig_mpodSketch1}}{\includegraphics[width=0.45\textwidth]{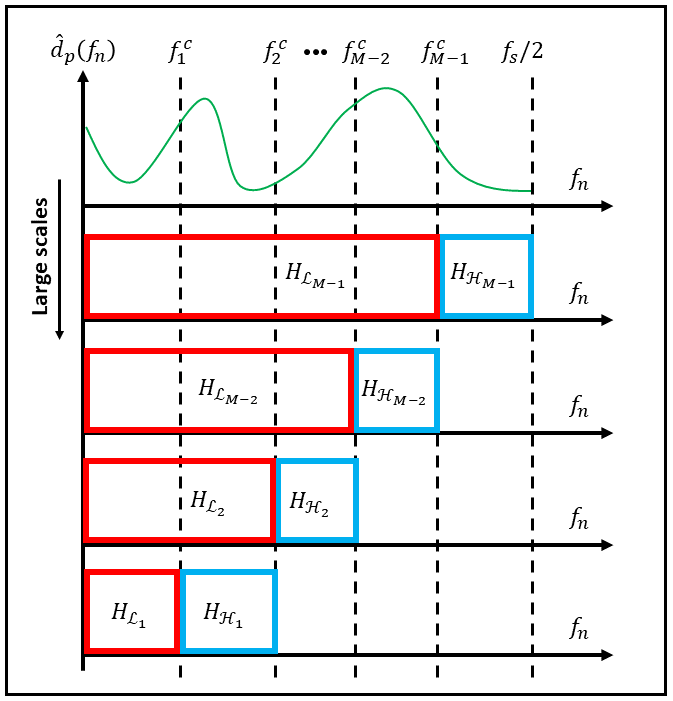}}\\
 	\centering
	\subfloat[]{\label{fig_mpodSketch2}}{\includegraphics[width=10.0cm]{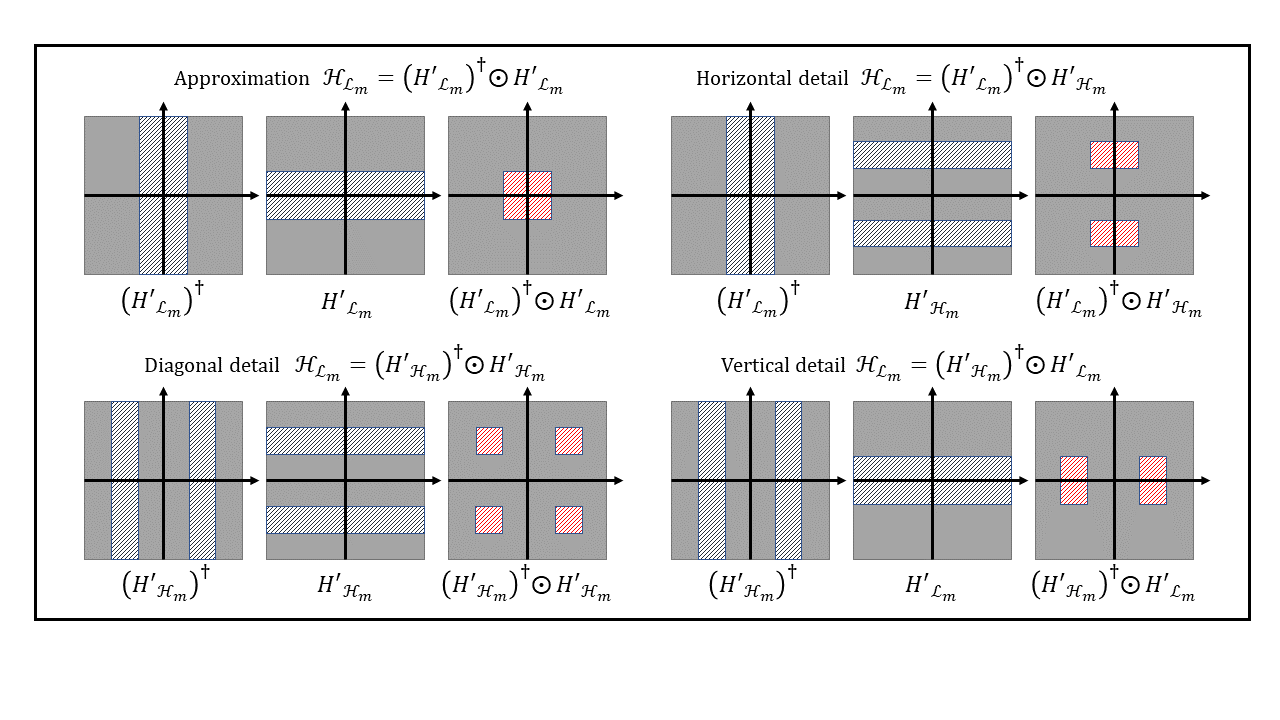}}
	\caption{Schematic depiction of the \mPOD~algorithm:
	in~\ref{fig_mpodSketch1}, the multi-resolution principle enacted by the filter bank; 
	in~\ref{fig_mpodSketch2}, an illustration of the different scales that can be \textit{revealed}.}
\end{figure}

The \mPOD~technique separates the scales in Eq.~\eqref{eq_fTTempCorrMat} by the application of 
a \textit{filter bank} $H_{\mathcal{L}_1},\ldots ,H_{\mathcal{H}_M}$ (see Fig.~\ref{fig_mpodSketch1})
to the Fourier--transformed temporal correlation matrix:
\begin{equation}\label{eq_fTTempCorrMat}
\hat{\mathbf{K}}=\mathcal{F}^{H}\,\mathbf{K}\,\mathcal{F}=\mathcal{F}^{H}\,(\mathbf{V}_1^{n_t})^T\,\mathbf{V}_1^{n_t}\,\mathcal{F}.
\end{equation}

Fig.~\ref{fig_mpodSketch1}  presents a sketch of the procedure (see Ref.~\cite{mendezEtAlJFM2019} for complete details).
In this manner, the \mPOD~method approximates the temporal correlation matrix as the superposition of $M$ different contributions (see Fig.~\ref{fig_mpodSketch2}):
\begin{eqnarray}\label{multiresolutionPOD}
\mathbf{K} \approx \mathbf{K}_{\mathcal{L}_1} + \sum\limits_{j=1}^{M-1}\mathbf{K}_{\mathcal{H}_j}.
\end{eqnarray}
\noindent Each of these contributions is, in turn, eigenfactored as $\mathbf{K}_{j}\,\mathcal{X}_j = \mathcal{X}_j\,\mathbf{D}_{s^2_{j}}$, \cfr~Eq.~\eqref{eq_pod_strategy2}. 
Next, the eigenvectors from the different scales are gathered into 
$\mathcal{X}_{mPOD}^0=[\mathcal{X}_{\mathcal{L}_1}, \mathcal{X}_{\mathcal{H}_{1}}, \ldots, \mathcal{X}_{\mathcal{H}_{m-1}}]$.
A \QR~factorization step is leveraged to correct eventual losses of orthogonality. Inverting the $\mathcal{R}$ factor leads to the right singular vector matrix $\mathbf{R}$:
\begin{eqnarray}
\mathcal{X}_{mPOD}^0\overset{QR}{=}\mathcal{Q}^{\perp}\,\mathcal{R} \Rightarrow \mathbf{R}\equiv \mathcal{Q}^{\perp} = \mathcal{X}_{mPOD}^0\,\mathcal{R}^{-1},
\end{eqnarray}
which are now guaranteed to be orthogonal.
Finally, the spatial modes (the left singular vectors) are obtained as $\mathbf{L}=\mathbf{K}\,\mathbf{R}\,\mathbf{D}_{s^2}^{-1/2}$, \cfr, Eq.~\ref{eq_pod_strategy2_modes}.

\begin{figure}[h]
	\centering
	\includegraphics[width=\textwidth]{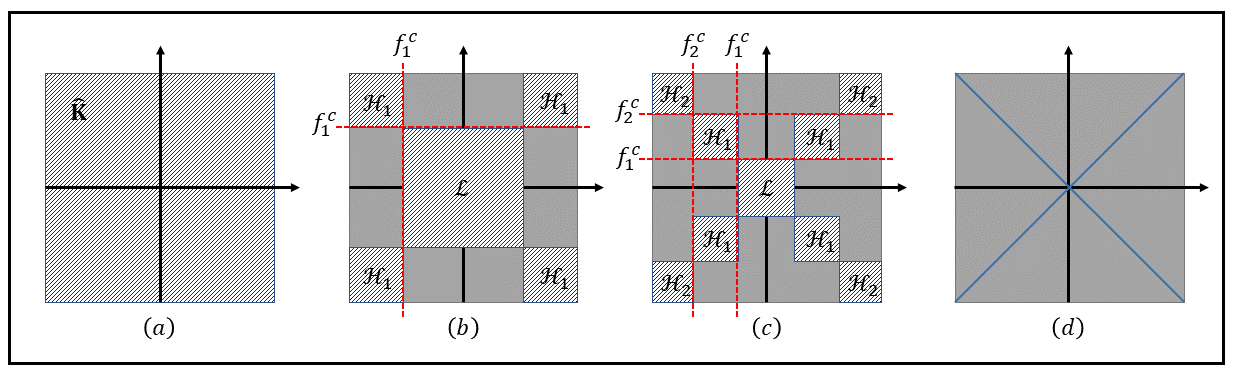}
	\caption{Schematic depiction of the partition of the correlation matrix $\hat{\mathbf{K}}$ by choosing different splitting frequencies $(f_i^c)$ vector. Case (a) considers a single scale, in which case \mPOD~becomes \POD. Cases (b) and (c) consider two and three scales, respectively. Case (d) considers $n_t/2$ scales; in this limiting case \mPOD~becomes \DFT.}\label{fig_mpodSketch3}
\end{figure}

The interested reader will find additional details on the \mPOD~method (\eg, how the filter bank is prepared) in the seminal publication~\cite{mendezEtAlJFM2019}.
One interesting property of the \mPOD~method is how the filter bank configuration defines the hierarchy of the scales resolved, Fig.~\ref{fig_mpodSketch3} illustrates this point.
The relationship of the \mPOD with the \DMD~and \RA~methods is also covered in Ref.~\cite{mendezEtAlJFM2019}.

\subsubsection{Multi-Resolution Dynamic Mode Decomposition}\label{subsec_MRDMD}

\begin{figure}[h]
	\centering
	\includegraphics[height=7.0cm]{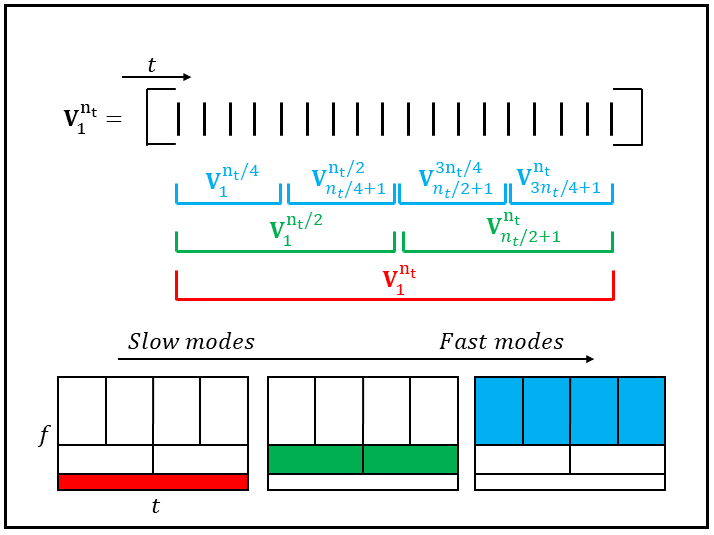}
	\caption{Schematic depiction of the method described in \cite{kutz2016multiresolution} for estimating \mrDMD~modes.}\label{fig_mrdmd_algorithm}
\end{figure}

Multi-resolution strategies can also be combined with the Dynamic Mode Decomposition method:
Kutz \etal, in Ref.~\cite{kutz2016multiresolution}, proposed a reformulation of the classical \DMD~method capable of 
segregating the data into a hierarchy of multiresolution time-scale components resembling that given in Eq.~\eqref{multiresolutionPOD}.
This reformulation, which is known as multi-resolution \DMD~(or \mrDMD), operates as indicated in Fig.~\ref{fig_mrdmd_algorithm}:
if $n_L$ levels are to be considered, 
the classical \DMD~method of \S~\ref{subsec_basicDMD} is applied $n_b\equiv2^{n_L-1}$ times to the recursively halved dataset (\ie, of length $n_t/2^l$), with $l=0,\ldots,n_L-1$.
At each level, only the $r_l$ \textit{slowest} (according to a user-specified threshold) modes are retained. 
The effect of this process is to recast Eq.~\eqref{eq_expansionDMD} as:
\begin{eqnarray}\label{eq_expansionDMD_mr}
  \mathbf{v}(t_j) = \sum_{i=1}^{r}\alpha_i \, \boldsymbol{\phi}_i \mu_i^{ j\, \Delta\,t^s} = 
  \sum_{l=1}^{n_L}\sum_{k=1}^{n_b}\sum_{i=1}^{r_l} \gamma^{l, k} \alpha_i^{l, k} \boldsymbol{\phi}_i^{l, k} (\mu_i^{l, k})^{ j\, \Delta\,t^s},
\end{eqnarray}
\noindent where $n_b$ and $r_l$ are respectively the number of bins and the number of \DMD~modes retained at level $l$,
and $\gamma^{l, k}$ is an \textit{indicator function} which equals $1$ if $t\in\left[t_{k},\,t_{k+1}\right]$ and $0$ elsewhere.



\subsection{Resolvent Analysis}\label{subsec_ra}
%
%
%
%
Let us now consider a linear(ized) dynamical system in continuous time subject to an external forcing:
\begin{equation}\label{eq_forcedlinearsystem}
\dot{\mathbf{v}}=\mathcal{A}^c\,\mathbf{v}+\mathbf{f}.
\end{equation}
In equation above, $\mathcal{A}^c$ is the Jacobian matrix of the system, 
and the state $\mathbf{v}$ may represent fluctuations about a statistically stationary state; 
non--linearities can be included into the external forcing $\mathbf{f}$.

It is well-known that the response  of the system to harmonic excitations $\mathbf{f}=\hat{\mathbf{f}}e^{-i\,\omega\,t}$ is, 
invoking the linearity of the system (see \eg~Ref.~\cite{bookBoyceDiPrima}):
\begin{equation}
\hat{\mathbf{v}}=\underbrace{(-\iu\,\omega \mathbf{I}-\mathcal{A}^c)^{-1}}_{\mathbf{H}(\omega)}\,\hat{\mathbf{f}}.
\end{equation}
\noindent The spectral properties of the \textit{Resolvent} operator $\mathbf{H}(\omega)$ defined above 
controls the ultimate fate of $\mathbf{v}$, see~\eg~Ref.~\cite{mckeonSharmaJFM2010}.
Resolvent Analysis~(\RA) techniques are covered in Ref.~\cite{jovanovicAnnualReview}, 
including an interpretation of Eq.~\eqref{eq_forcedlinearsystem} as the state--transition equation in a state--space representation of the system~(\ie, input--output analysis, see also Refs.~\cite{mckeonSharmaJFM2010,jovanovicAnnualReview}). 
Additionally, Towne~\etal~reveal in Ref.~\cite{towneEtAlSPOD} the connections between the \SPOD~algorithm and~\RA~techniques.
However, and for brevity, in this work we focus exclusively on the data--driven formulation of the Resolvent Analysis (\RA) technique recently proposed by Herrmann \etal~in Ref.~\cite{dmdResolventJFM}, which we discuss briefly next.

The data--driven Resolvent Analysis in Ref.~\cite{dmdResolventJFM} resorts to the exact \DMD~algorithm~\cite{tuRowleyEtAlOnDMDTheoryApplications}.
More specifically, if a certain weight matrix $\mathbf{W}$ is considered, 
first the transformed data sequence $\mathbf{U}_1^{n_t}=\mathbf{Q}\,\mathbf{V}_1^{n_t}$ is built;
$\mathbf{Q}$ is the Cholesky factor of the weight matrix, namely $\mathbf{W}=\mathbf{Q}^T \, \mathbf{Q}$.
Then,  partial subsequences $\mathbf{X}\equiv\mathbf{U}_1^{n_t-1}$ and $\mathbf{Y}\equiv\mathbf{U}_2^{n_t}$ 
are related through a variant of Eq.~\eqref{eq_dmd_3} (see also Ref.~\cite{liEtAl2022ThetaDMD}), \ie:
\begin{eqnarray}
\mathbf{Y}=\underbrace{exp\left(\mathcal{A}^{c}\,\Delta t^s\right)}_{\mathbf{B}}\;\mathbf{X}.
\end{eqnarray}
\noindent Considering then a $r'$-rank  \SVD--reduced matrix $\bar{\mathbf{B}}$, 
its eigenvalues $\boldsymbol{\Lambda}_{\bar{B}}$ and  
direct ($\boldsymbol{\Psi}_{\bar{B}}$) and  adjoint ($\boldsymbol{\Psi}_{\bar{B}}^{\dagger}$) eigenvectors are  computed.
From them, an intermediate matrix and its \QR~decomposition:
$\boldsymbol{\Psi}_{\bar{B}}^{H}\;\mathbf{W}\;\boldsymbol{\Psi}_{\bar{B}}\overset{\QR}{=}\widetilde{\mathbf{F}}^{H}\widetilde{\mathbf{F}}$ 
\noindent leads to the approximated resolvent operator:
\begin{eqnarray}\label{eq_resolvent_svd}
\widetilde{\mathbf{F}}\;\left(-\omega\,\mathbf{I} - \boldsymbol{\Lambda}_{\bar{B}} \right)\overset{\SVD}{=}
\mathbf{L}_{\bar{B}}\;\mathbf{S}_{\bar{B}}\;\mathbf{R}_{\bar{B}}^{H}.
\end{eqnarray}
\noindent  whose  complex  \SVD~allows to obtain the \textit{forcing} modes as:
\begin{eqnarray}\label{eq_raDD_forcingmodes}
\boldsymbol{\Phi}= \boldsymbol{\Psi}_{\bar{B}}\;\widetilde{\mathbf{F}}^{-1}\;\mathbf{R}_{\bar{B}},
\end{eqnarray}
whereas the \textit{response} modes are given by:
\begin{eqnarray}\label{eq_raDD_responsemodes}
\boldsymbol{\Psi}= \boldsymbol{\Psi}_{\bar{B}}\;\widetilde{\mathbf{F}}^{-1}\;\mathbf{L}_{\bar{B}}.
\end{eqnarray}

\clearpage
\newpage

\section{Numerical databases}\label{sec:NumSim}



In this work we aim at evaluating the performance of several publicly available data--driven, modal decomposition feature identification strategies. 
In an attempt to conduct an informative as possible  comparison we will consider three different, increasingly complex flow datasets. 
The datasets chosen, which are also publicly available, are representative of three distinct flow regimes: laminar, turbulent and transitional.

The first testcase considered  (which we term TC1) is the incompressible flow field around the mid-section of a very long cylinder. 
The flow dynamics of a relatively low Reynolds flow past a circular cylinder is well known,  and has been studied in detail by Barkley \& Henderson \cite{BarkleyHenderson96}, among others. 
It is known that the flow bifurcation process in the wake of a circular cylinder starts at $\Rey_D\simeq48$, where a Hopf bifurcation triggers the flow transition from symmetric, steady flow to unsteady two-dimensional flow~\cite{Jackson87}.
Specifically,  we consider a $\Rey_D=U\,D/\nu=100$ flow, with $U$ the free stream velocity, $\nu$ the kinematic viscosity and $D$ the diameter of the cylinder.
At these conditions the flow is laminar and two-dimensional. The flow shows a periodic vortex shedding with leading non-dimensional frequency Strouhal number $\St=f\,D/U=0.16$, with $f$ the frequency in Hertz. 
All in all, this dataset is an ideal testcase to evaluate the performance of the methods considered herein.

\begin{figure}[h]
	\centering
	\includegraphics[width=6cm]{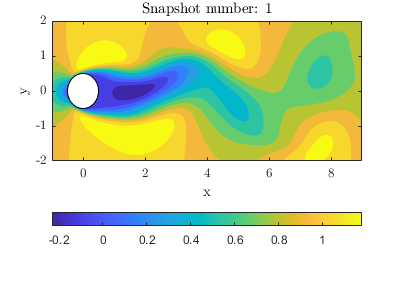}
	\includegraphics[width=6cm]{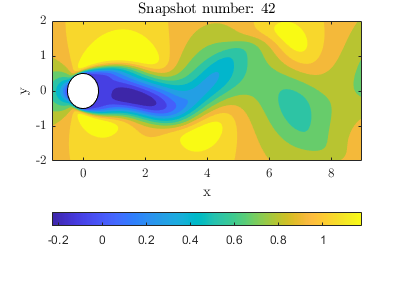}\\
	\includegraphics[width=6cm]{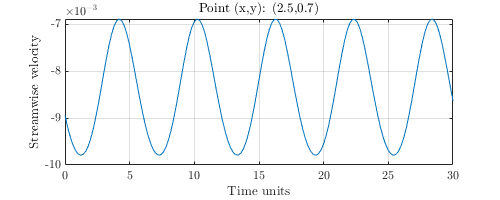}
	\includegraphics[width=6cm]{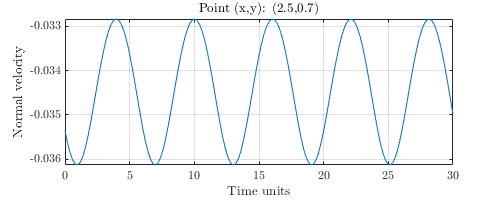}
	\caption{Flow past a circular cylinder at $\Rey_D=100$. 
	Top: two representative snapshots of the streamwise velocity field. 
	Bottom: temporal evolution of the streamwise (left) and normal (right) velocity components
	at a tracer point $(2.5,0.7)$ located in the wake. 
	}\label{fig:cyl2D}	
\end{figure}

The database employed is taken from Ref.~\cite{kutzEtAlBook}. 
The characteristics of this dataset (size per snapshot $n_p$, number of snapshots $n_t$, sampling period $\Delta\,t^s$, most representative Strouhal,  \ldots) is summarized in Table~\ref{tab:summaryDataBase}.
Figure \ref{fig:cyl2D} shows two representative snapshots of the stramwise velocity component in the database analysed. The figure also shows the temporal evolution of the streamwise and normal velocity components in a tracer point located at the wake: the periodic character of the velocity is evident.

The second testcase (TC2), originally described in Ref.~\cite{towneEtAlSPOD}, considers 
the pressure field in a turbulent jet flow.
Although it is notorious that, in contrast to wall bounded turbulent flow \cite{leClaincheHODMDChannel}, the frequency spectrum in jet flows presents selected high-amplitude frequencies, which are in charge of driving the flow dynamics. These high-amplitude frequencies are in many cases connected with flow instabilities occurring in the shear layer of the jet, which is in good agreement with the own nature of this type of flows.

\begin{figure}[h]
   	\centering
	\includegraphics[width=6cm]{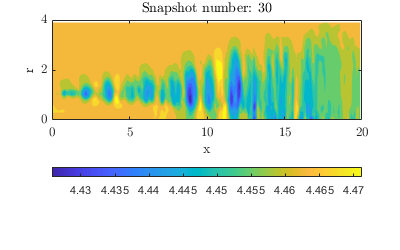}
	\includegraphics[width=6cm]{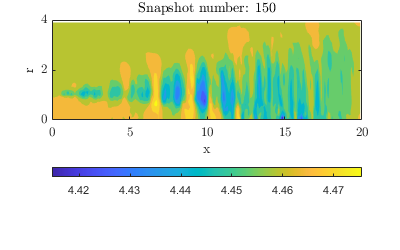}\\
	\includegraphics[width=6cm]{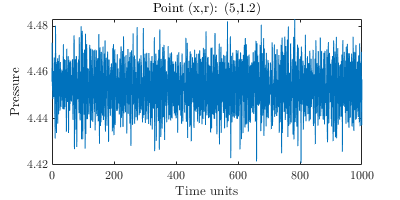}
	\includegraphics[width=6cm]{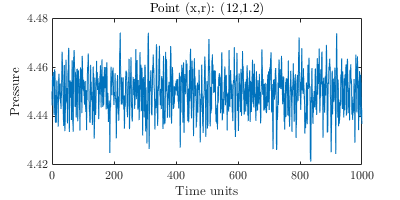}
	\caption{Turbulent jet flow at $\Rey_D=10^{6}$ and $\M=0.4$. 
	Top: two representative snapshots of the pressure field. 
	Bottom: temporal evolution of the pressure at two representative tracer points, 
	located at $(x,r)=(5,1.2)$ (left) and  $(x,r)=(12,1.2)$ (right).}\label{fig:jet}	
\end{figure}

Specifically, testcase TC2 is a turbulent jet at $\Rey_D=\rho_j\,U_j\,D/\mu_j\approx10^{6}$, Mach number $\M=U_j/c_j=0.4$ and $T_j/T_\infty=1$.
Here $U$ is the velocity, $c$ is the speed of sound, $T$ is the temperature, $\rho$ is the density, D is the nozzle diameter, $\mu$ is the dynamic viscosity;  the subscripts $j$ and $\infty$ refer respectively to the mean conditions at the nozzle exit and  at the  far field.
At the present flow conditions, the flow is characterized by a high-amplitude frequency driving the flow dynamics at  $\St\simeq0.6$  (more details about this flow characterization in Ref. \cite{towneEtAlSPOD}).
The high complexity and the turbulent character of this database is illustrated in Fig. \ref{fig:jet}:
differences with Fig.~\ref{fig:cyl2D}) are striking, specially evident for the pressure signal behaviour.

This dataset has been derived from a Large Eddy Simulation computation, and hence contains a wide range of spatio-temporal scales interacting in complex manners. 
The dataset thus poses a  challenge to the data analysis methods considered in this work. 
As before, the characteristics of this dataset are summarized in Table~\ref{tab:summaryDataBase}.

The third testcase (TC3) considers yet another incompressible laminar cylinder flow,
but this time at $\Rey_D=280$, which is significantly more complex than TC1.
As established above, beyond the first bifurcation at $\Rey_D\simeq48$, 
the wake after the circular cylinder becomes unsteady, but the flow remains two--dimensional.
If $\Rey_D$ is further increased, a second flow bifurcation emerges at $\Rey_D\simeq 190$, 
and the bidimensional flow evolves into a three-dimensional flow \cite{BarkleyHenderson96}.
Depending on the spanwise length of the cylinder, defined as $L_z=2\pi/\beta$ with $\beta$ the spanwise wavenumber, differences are found on the value for the critical Reynolds number and the system dynamics \cite{BlackburnEtAl2005}. 

In this work, we consider $L_z=6.99$, which corresponds to a periodic flow, where a three-dimensional instability is leading the flow changes driven by a mode with critical wavenumber $\beta\simeq 7.570$ (more details in Ref. \cite{LeClaincheetalFDR18}). 
Thus, at the present flow condition (\ie~$\Rey_D=280$, $L_z=6.99$), the stationary solution is periodic, with leading frequency $\St=0.21$. 
In the transient regime of the simulation, frequency $\St=0.36$ is also identified with high amplitude but negative growth rate, as we shall discuss below.
The latter frequency is connected to the transition to a three--dimensional quasi--periodic flow, 
which is a flow bifurcation that takes place at $\Rey_D \simeq 380$, see Blackburn \etal~\cite{BlackburnEtAl2005}. 
Note that this second high-amplitude frequency disappears when the analysis is carried out in the saturated flow (\ie, when the stationary state is reached). 

\begin{figure}[h]
	\centering
	\includegraphics[width=6cm]{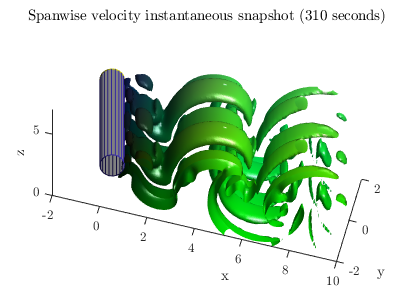}
	\includegraphics[width=6cm]{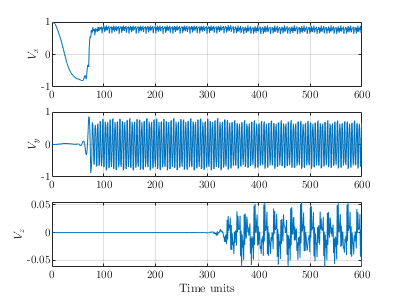}
	\caption{Three-dimensional wake visualization of a circular cylinder flow at $\Rey_D=280$. 
	Left: spanwise velocity calculated at a representative snapshot from the transient regime. 
	Right: temporal evolution of the streamwise (top), normal (center) and spanwise (bottom) velocity components at tracer point $(5,0,0.83)$ 
    }\label{fig:cyl3D}	
\end{figure}

Dataset TC3, available from Ref.~\cite{bookLeClaincheHODMD}, contains unsteady three-dimensional flow data (see Fig.~\ref{fig:cyl3D}).
The database includes snapshots collected since the beginning of the numerical simulation,
which allows to assess the performance of the data--driven methods also for transient solutions,
thus revealing the capabilities and limitations of these tools when predicting the leading dynamics of the flow. 
In other words, TC3 illustrates the performance of the methods as predictive reduced order models. 
If the dynamics are properly identified in transient solutions, 
then it is possible to reduce the computational cost of numerical simulations, which could be suitable for future flow control applications. 
As before, the characteristics of the database are summarized in Table~\ref{tab:summaryDataBase}.




\begin{table}[h]
    \centering
    \caption{ Summary of the main characteristics of the three testcases. $n_p$ and $n_t$ represent the dimensions of the snapshot matrix (Eq. (\ref{eq_dataMatrixDefintion})) on the $X$ and $Y$ axis, respectively and $\Delta t^s$ represent the sampling rate. 
}\label{tab:summaryDataBase}

    \makebox[1.0\textwidth]{\makebox[1.35\textwidth]{%
    \centering
    {\normalsize
        \resizebox{1.35\textwidth}{!}{%
		\begin{tabular}{l | c c c c |  c  c | c}
		    \hhline{-|----|--|-}
			Testcase          & {\footnotesize$[x_{min},x_{max}]\times[y_{min},y_{max}]\times[z_{min},z_{max}]$}& $n_x\times n_y \times n_z$&  Variables&  $n_p$ & $n_t$   &  $\Delta t^s$   & Target $\St$ \\
 			\hhline{-|----|--|-}
			TC1  (laminar, 2D)            &       $[-1,9]\times[0,2]$         &        $449\times199$&      $(u_x, u_y)$&  $1.78\times10^5$&   $151$&   $0.2$&  $0.16$ \\
			TC2  (turbulent, axisymmetric)&      $[0, 20]\times[0,4]$         &         $39\times175$&               $p$&  $6.83\times10^3$& $5000$&   $0.2$&   $0.6$\\
			TC3  (transitional, 3D)       & $[0,10]\times[-2,2]\times[0,6.99]$& $100\times40\times64$& $(u_x, u_y, u_z)$&  $7.68\times10^5$& $599$&     $1$&  $0.21$ ($0.36$) \\
 			\hhline{-|----|--|-}
		\end{tabular}
        }
	}
    }}

\end{table}

\section{Method comparison}\label{sec_methodComparison}

Section~\S~\ref{sec_methodology} has described in, hopefully, sufficient detail the different methods considered.  In This section we offers an additional comparison where we investigate whether the different techniques considered are capable of 
identifying the most relevant frequencies: $St\approx 0.16$ for TC1, $St\approx 0.6$ for TC2 and $St\approx 0.21$ and $St\approx 0.36$ for TC3,  as the \textit{quality} of the dataset is worsened by shortening (\ie, by discarding an increasing number of the latter temporal samples of the dataset). This test assesses how good the different techniques are in extracting meaningful information from partial data.\\

\begin{figure}[H]
     \subfloat[\DMD, \HODMD, \mrDMD, and \RA. \label{fig_Cyl2D_St_1}]{\includegraphics[width=1.0\textwidth]{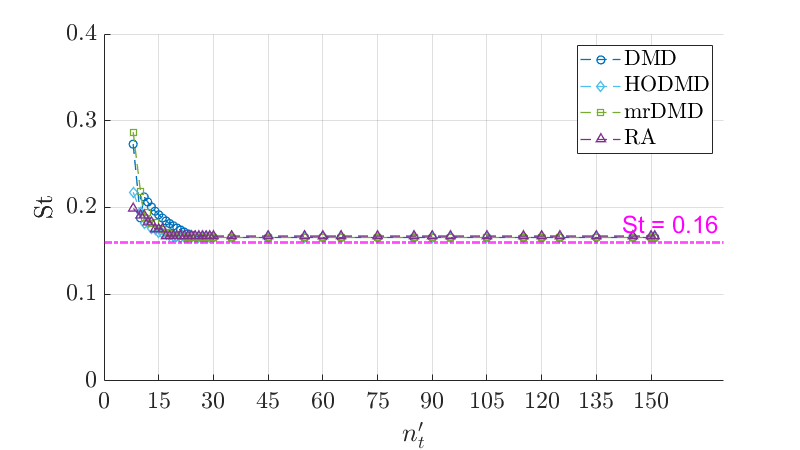}}\\
     \subfloat[\POD, \mPOD, \SPOD, and \FFT. \label{fig_Cyl2D_St_2} ]{\includegraphics[width=1.0\textwidth]{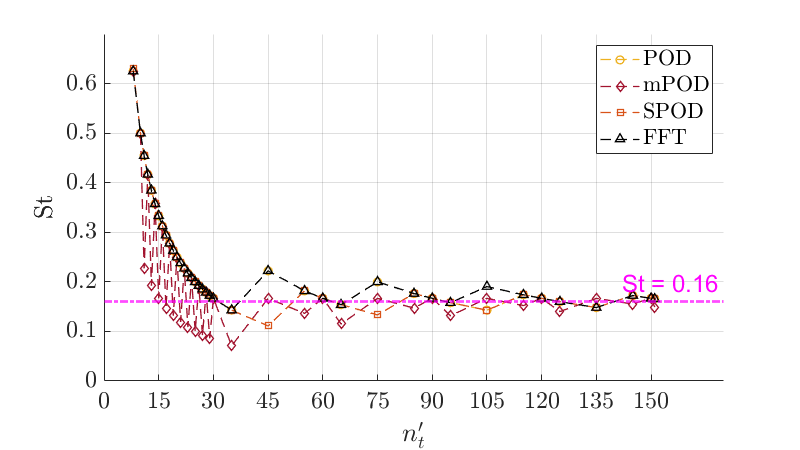}}
     \caption{
     TC1: dominant $\St$ number detected \Vs~dataset temporal length.}\label{fig_2dCyl_st}
 \end{figure}

\begin{figure}[H]
     \subfloat[\DMD, \HODMD, \mrDMD, and \RA. \label{fig_jet_St_1}]{\includegraphics[width=1.0\textwidth]{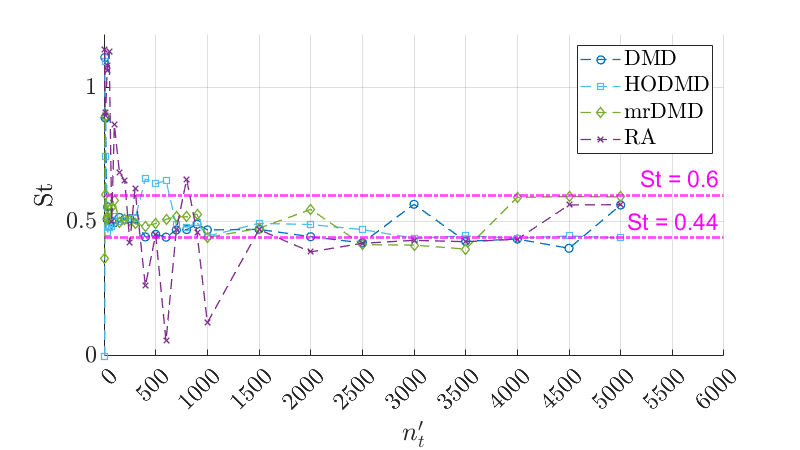}}\\
     \subfloat[\POD, \mPOD, \SPOD, and \FFT. \label{fig_jet_St_2} ]{\includegraphics[width=1.0\textwidth]{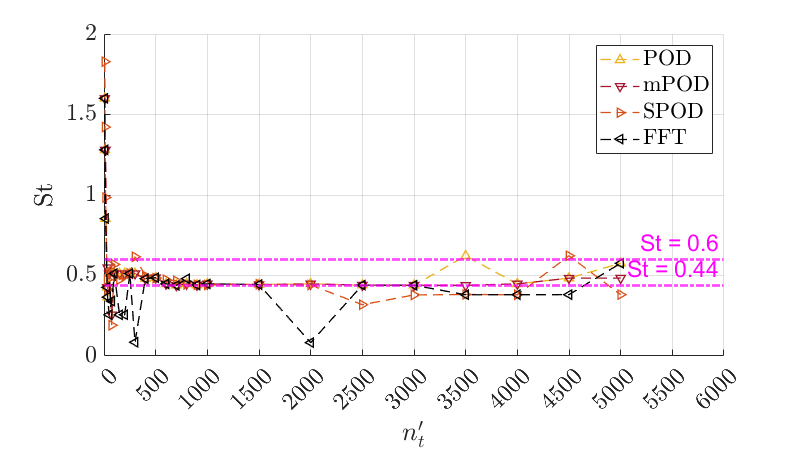}}
     \caption{TC2: dominant $\St$ number detected \Vs~dataset temporal length.}\label{fig_jetTurb_st}
 \end{figure}

\begin{figure}[H]
     \subfloat[\DMD, \HODMD, \mrDMD, and \RA. \label{fig_Cyl3D_St_1}]{\includegraphics[width=1.0\textwidth]{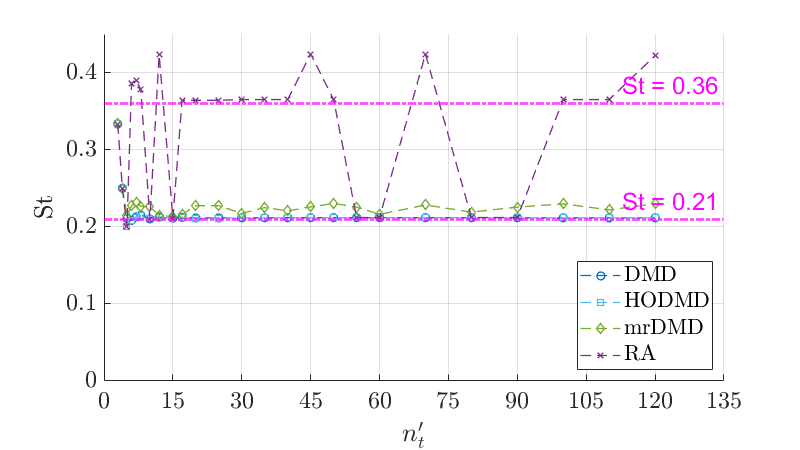}}\\
     \subfloat[\POD, \mPOD, \SPOD, and \FFT. \label{fig_Cyl3D_St_2} ]{\includegraphics[width=1.0\textwidth]{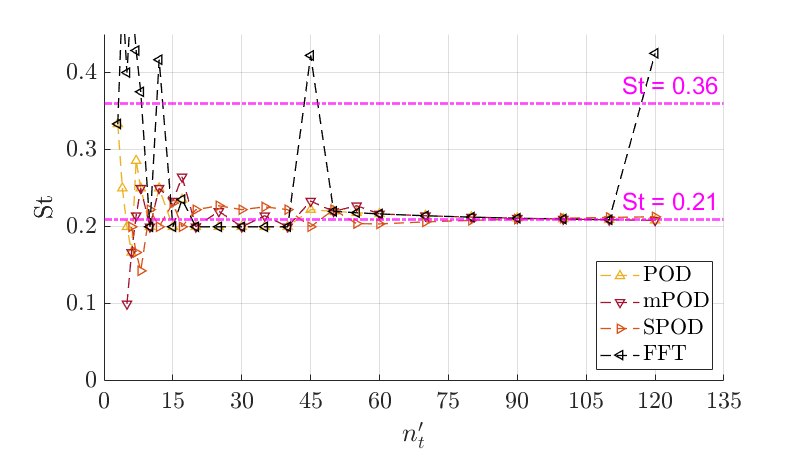}}
     \caption{TC3: dominant $\St$ number detected \Vs~dataset temporal length.}\label{fig_3dCyl_st}
 \end{figure}

The results of the aforementioned evaluation are presented on a testcases basis. We start by considering the results for TC1, as can be seen in Fig. \ref{fig_Cyl2D_St_1}, \DMD~and \DMD-based methods functions very well in identifying the dominant frequency, even when the data is limited (shortened). However, if we take a look to Fig. \ref{fig_Cyl2D_St_2}, we can see that \FFT~and \POD-based techniques perform poorly when compared with \DMD-based methods. If we consider \FFT~as an example, we can see that this technique only identifies the right frequencies when the number of snapshots taken is a multiple  of 30, otherwise the technique is experiencing what is called "spectral leaking", which is the same case as \POD, \SPOD~or \mPOD. Hence these techniques may perform sub-optimally in cases sparse data, unlike \DMD~based approaches. Also, to obtain similar results, \FFT~requires a longer data sequence than the other methods, which can be a disadvantage for large databases. However, \FFT~is a robust method that always identifies the relevant dynamics, even when the spectral complexity is much larger as the spatial complexity, which is a common case in the performance of experimental sensor measurements. In TC2, The dataset describes a complex, turbulent pressure field. As it is usual in turbulent flows,  a large number of temporal samples is necessary to provide a good representation of the underlying physical process. In this manner, the initial number of samples ($n'_t=5000$) justifies why the results  provided by the different methods are reasonably good, as we can see in Figures~\ref{fig_jet_St_1} and~\ref{fig_jet_St_2}.
The large number of spatio-temporal scales driving the dynamics of turbulent flows, makes it challenging to identify the most relevant dynamics. However, the present test case studied the main changes in the flow are driven by a strong global flow instability, that all the methods are capable to identify. The definition of the rest of the dynamics varies depending of the method used, consequently varying the reconstruction error of the main database. In TC3, the methods are capable to capture the main dynamics, even in the transient regime. Similarly to TC2, the flow instabilities are well defined. The method differences are found in the accuracy of the main frequencies captured and the reconstruction of the main dynamics. (as shown in Figures~\ref{fig_Cyl3D_St_1} and~\ref{fig_Cyl3D_St_2}).\\
Hence, regarding the achievement of the main goal, which is capturing the dominant frequencies: $St\approx 0.16$ for TC1, $St\approx 0.6$ for TC2 and $St\approx 0.21$ and $St\approx 0.36$ for TC3, all the methods were successful, especially in TC1. In the more complex cases, as in TC2 and TC3, we can notice a small fluctuation in the frequencies depending on the number of snapshots, where some methods may suffer spectral leaking and others may identify several frequencies with comparable amplitude levels. However, these results are considered reasonable because of the flow complexity, which is larger in the turbulent and transient flow. Consequently, the methods are expected to face difficulties in identifying the dominant dynamics and avoiding the components related to the small flow scales of the transient regime. Nevertheless, all the methods eventually converge and we can highlight the fact that the \HODMD~algorithm is the fastest to converge in all testcases.

 \clearpage
 \newpage
 
In the following of the paper, we will try to understand the performance of all the methods studying carefully each one of the testcases.

\section{Results}\label{sec:Results}
In this section we discuss the results obtained when applying the different data--driven modal decomposition methods to the datasets TC1--TC3.
We follow the structure laid out in section~\ref{sec_methodology}, and begin with the results obtained with the most classical methods (\POD, \DMD~and \FFT), and continue  with the improved methods that employ data redundacy (\SPOD~and \HODMD).
We follow next with multi--resolution methods (\mPOD~and \mrDMD), and finish with data--driven resolvent analysis (\RA).

All the computations have been performed on a desktop computer equipped with a $4$-core Intel$(R)$ Core$(TM)$ \texttt{i5-3570K} CPU at $3.40\,GHz$, a~cache memory of $6144\,kB$ and $8.0\,GB$ of \texttt{RAM}. 


\subsection{Feature detection using classical methods: \POD, \DMD~ and \FFT\label{sec:ResClassic}}

\begin{figure}[h]
	\centering
	\includegraphics[width=6cm]{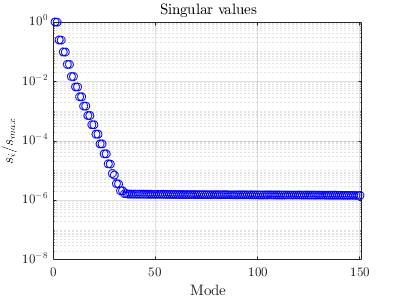}
	\includegraphics[width=6cm]{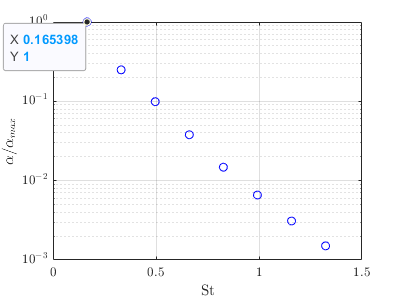}
	\caption{TC1. Left: singular values from the \POD~ analysis. Right: \DMD~spectrum comparing non-dimensional frequencies with the \DMD~amplitudes, using $\varepsilon=1\cdot 10^{-3}$ in Eq.~\eqref{eq_truncationStrategy}.}\label{fig:SVcyl2D}	
\end{figure}

This section describes the results obtained through application of the most established methods, namely \POD, \DMD~and \FFT~to the datasets TC1-TC3.
Figure \ref{fig:SVcyl2D}-left shows the singular values obtained when \POD~is applied to analyse TC1, the two-dimensional cylinder wake.
Observe how, the singular values are grouped in pairs of frequencies, which is a reflection of the strong periodic character of the flow \cite{OberleithnerJFM}. 
Fig.~\ref{fig:SVcyl2D}-right shows the amplitudes of the \DMD~modes as function of frequency $\St$, which, by a certain abuse of language, is usually termed \DMD~\textit{spectrum}. 
The spectrum shows that the most relevant mode appears at the leading frequency $\St\approx0.16$; 
$7$ distinctly large amplitudes also appear at integer multiples of the leading frequency: these are harmonics of the leading mode.
Note that in the stationary regime, the growth rates $Re(\lambda)$, should be zero. 
In noisy databases (or numerical simulations that still are not saturated in time), 
the growth rates are never exactly zero, since the value they take is of the same order of magnitude as the level of noise. 
This in turn provides an estimation of the error made in the frequency calculations (see more details in Refs. \cite{DukeSoriaHonnery2012,LeClaincheVegaSoria17,LeClaincheVegaPoF17}).



\begin{figure}[H]
	\centering
	\includegraphics[width=6cm]{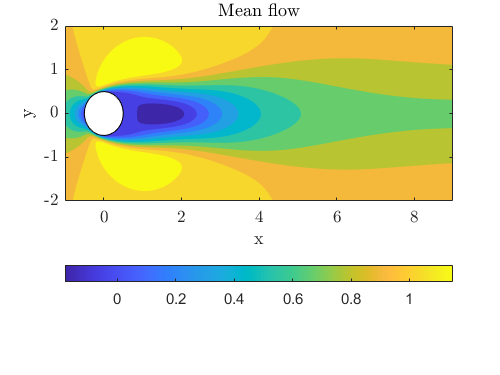}
	\includegraphics[width=6cm]{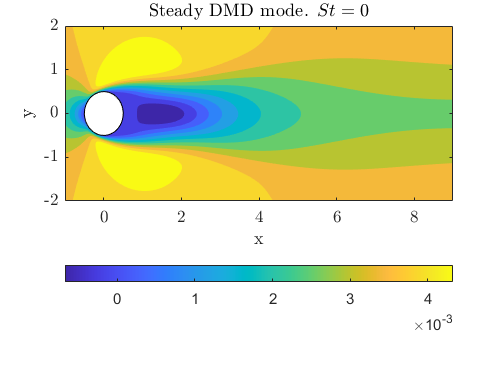}\\
    \includegraphics[width=6cm]{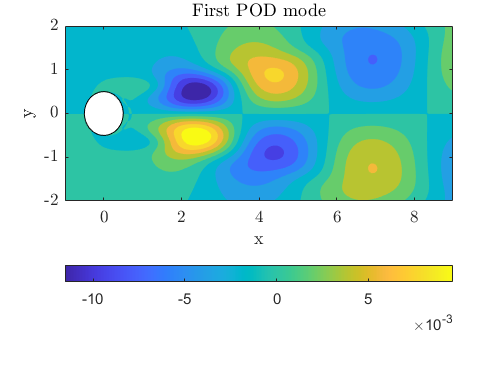}
    \includegraphics[width=6cm]{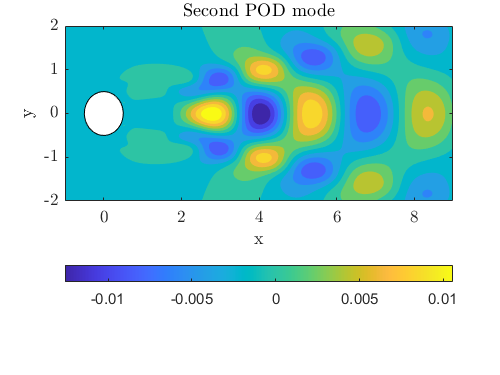}\\
    \includegraphics[width=6cm]{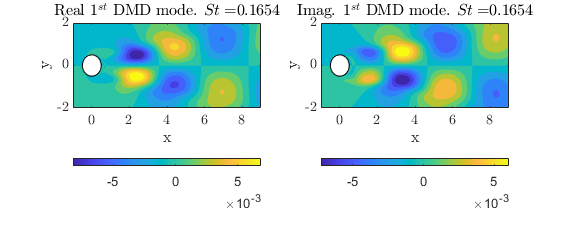}
	\includegraphics[width=6cm]{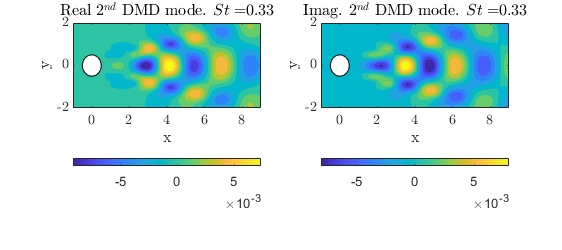}\\
	\caption{TC1: mean flow (top-left) and steady ($\St=0$) \DMD~mode (top-right). 
	\POD~(middle) and \DMD~(bottom) dominant modes; from left to right: first and second highest energy/amplitude modes (\DMD~modal frequencies: $\St=0.1654,0.33$).
    }\label{fig:PODmodesDMDmodes2DCyl}	
\end{figure}

Fig. \ref{fig:PODmodesDMDmodes2DCyl}~shows the mean flow and the most relevant \POD/\DMD~modes.
Note how the mean flow and the \DMD~mode at $\St=0$ are similar, suggesting that the steady \DMD~mode represents the mean flow, see Ref. \cite{LeClaincheetalFDR18}. 
Regarding the remaining modes, observe how these modes are qualitatively similar. Recall that (Sec~\ref{sec_methodology}) \POD~modes are real, whereas \DMD~modes appear in complex conjugate pairs; note also how the real and imaginary part of the \DMD~modes are phase shifted, revealing the travelling character of the modes in the wake flow. 


\begin{figure}[h]
	\centering
	\includegraphics[width=12cm]{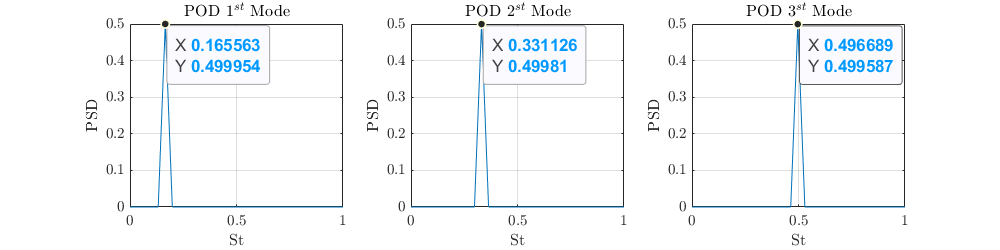}
	\caption{TC1: frequency contents corresponding to \POD~modes, obtained from the chronos matrix $\mathbf{R}_0$, Eq.~\eqref{eq_spatialReduction0}. 
	From left to right: first, second and third highest energy modes.}\label{fig:PODspectrum2DCyl}	
\end{figure}

However, note that TC1 is a dataset describing a two-dimensional, laminar, coherent and well organized motion (the wake) generated by a highly accurate numerical solver (\ie, ~relatively \textit{clean} from noise): it should not be surprising that \POD~and \DMD~analyses provide comparable results.
For example, Fig.~\ref{fig:PODspectrum2DCyl} shows the frequency content of the first three dominant \POD~modes (calculated using \FFT~on the corresponding temporal \POD~mode $\mathbf{r}_{0,j}$, Eq.~\eqref{eq_spatialReduction0}). This behaviour is tipically observed whenever both \DMD~and \POD~are applied to analyse transient data from linear solutions. In those cases, \POD, \DMD, but also linear global stability modes are the same (see details and more results in Ref.~\cite{Gomezetal12}).
%

In fluid dynamic applications, this \SVD-enabled reduction is useful at two levels: 
on the one hand, since most datasets are densely discretized in space and sampled frequently in time (\ie, they are large);
\SVD~helps to reduce the memory footprint of the \DMD~analysis. 
On the other hand, the optimal representation properties of \SVD~become useful to filter noise and spatial redundancies, \ie~ to \textit{clean} the data. 
At each \SVD~step\footnote{There is one \SVD~step for  classical \DMD, and two for \HODMD, section~\ref{sec_methodology}},
the rank of the scaled chronos matrix $\mathcal{C}_1^{n_t-1}$ (Eq.~\eqref{eq_hodmdFirstSVD}) can be controlled directly (setting $r'$) or indirectly (via the tolerance $\varepsilon_{1}$, Sec.~\ref{subsec_HODMD}). 
Usually, larger tolerances (\ie, more \SVD~modes retained) help the method to properly identify the main dynamics of the system. 
However, and somewhat contradictorily, complex flows datasets (\ie, multi-scale, turbulence, noise, etc.) ask for as intense dimensionality reductions as  possible. The  user needs thus to perform this reduction carefully, as suboptimal $\varepsilon_{1}$ values can lead even to failure of the method.
This behaviour is connected to the spectral and spatial complexities introduced in sections~\ref{subsec_basicDMD} and~\ref{subsec_HODMD}. Similarly, the \DMD~frequency spectrum is also sensitive to the number of grid points composing the spatial domain.  
For instance, in experimental measurements, if the number of spatial points (spatial complexity) is smaller than the spectral complexity (number of frequencies retained in the \DMD~expansion Eq. \eqref{eq_expansionDMD_mr}), then \DMD~fails. 
In such cases, other methods, like \POD, \FFT, or the more robust approaches \SPOD~and \HODMD~arise as sensible alternatives.

\begin{figure}[H]
	\centering
	 \subfloat[\label{fig:TC1_fft_spatial_NG_a} ] {\includegraphics[width=0.5\textwidth]{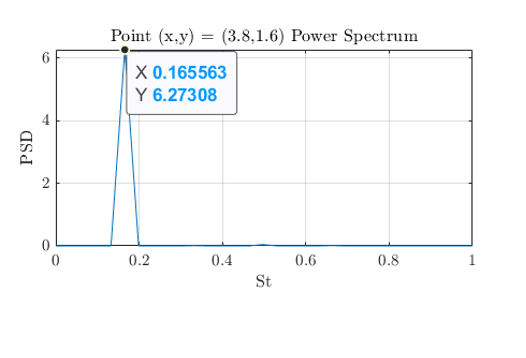}}
\subfloat[\label{fig:TC1_fft_spatial_NG_b} ] {\includegraphics[width=0.5\textwidth]{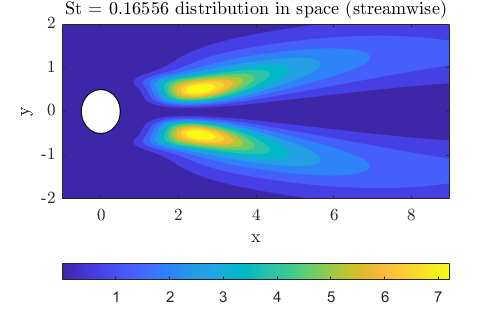}}\\	 

	\caption{TC1: Frequency spectrum, computed using Eq.~\eqref{eq_psd} (\ref{fig:TC1_fft_spatial_NG_a}) and spatial distribution at $\St\approx0.16$ (\ref{fig:TC1_fft_spatial_NG_b}) obtained with \FFT.}\label{fig:FFT2DCyl}	
\end{figure}

We present now the \FFT~results. Fig.~\ref{fig:FFT2DCyl}-top shows the Fourier spectrum calculated for TC1, computed using Eq.~\eqref{eq_psd}.
Observe how for this simple, strongly periodic flow, \FFT~identifies the same frequencies as \POD~and \DMD. 
Fig. \ref{fig:TC1_fft_spatial_NG_b} shows the spatial distribution of the horizontal component of the \FFT~mode at $\St\approx0.16$, 
This information reveals that most \textit{energetic} areas lie in the wake of the cylinder. 
According to the horizontal component, relevant information lies symmetrically above and below the $y=0$ line.
The spatial information provided by the \FFT~method is not as easily interpretable as that provided by the \POD/\DMD~methods.
However, this is easy to anticipate, if one recalls that \FFT~operates on the temporal correlation matrix (see Sections~\ref{subsec_FFT} and~\ref{subsec_MR}),
which means that spatial information is given up to a certain extent. 
%

As we have observed so far, classical data--driven modal decomposition methods, namely \POD, \DMD~and \FFT~provide comparable results whenever they are applied to flow databases describing a strongly coherent behaviour. 
We investigate now how these techniques perform when confronted with more complicated flows, namely the fully turbulent jet flow (TC2) and the transitional, three-dimensional cylinder flow (TC3) databases. 
In this last case, we consider two differentiated temporal ranges, which we respectively term \textit{transient regime} $\left (t\in[300,\,420]\right )$ and \textit{saturated regime} $\left(t\in[450,\,570]\right)$, see Fig.~\ref{fig:cyl3D}-right.

\begin{figure}[H]
	\centering
	\includegraphics[width=6cm]{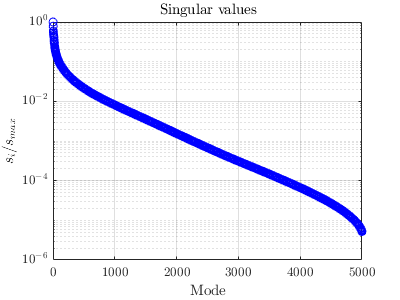}
	\includegraphics[width=6cm]{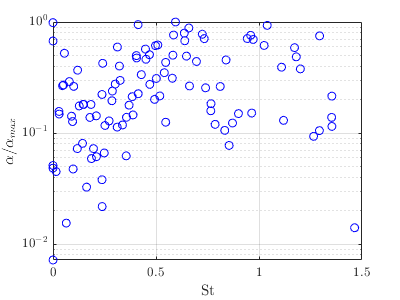}
	\caption{TC2. Left: singular values from the \POD~analysis. Right: \DMD~spectrum comparing non-dimensional frequencies with the \DMD~amplitudes, using $\varepsilon=2\cdot 10^{-1}$ in Eq.~\eqref{eq_truncationStrategy}.}\label{fig:SV_dmdAmp_TC2}	
\end{figure}

Concerning TC2, Figure~\ref{fig:SV_dmdAmp_TC2}-left presents the singular values obtained with \POD.
Observe the radical difference in the decay rate of the singular values for TC2 with respect to the more simple TC1: whereas for TC1 (Fig.~\ref{fig:SVcyl2D}-left) the $s_i$'s (Eq.~\eqref{eq_spatialReduction0}) decay relatively fast and stabilize at around $10^{-6}$,
the $s_i$'s for TC2 decay also very fast at first, and slightly slower afterwards;
however, there is not an index $i$ beyond which the singular values become  negligible, or even stabilize. 
This behaviour is consistent with the multiscale nature of turbulence:
in a turbulent flow, all the scales contribute with \textit{energy}.


Figure~\ref{fig:SV_dmdAmp_TC2}-right shows the \DMD~spectrum:
observe how relevant modes appear to be the steady component and that at $\St\approx0.6$ and $St\approx0.43$.
However, several \DMD~modes present amplitudes at relatively high levels, some of them comparable to the mode at $\St\approx0.6$.
And in this case it is much more difficult to establish a clear amplitude-based criterion to consider or not a specific \DMD~mode.
In this turbulent case, the overall picture is not as clear as it was for TC1.

\begin{figure}[H]
	\centering
	\includegraphics[width=6cm]{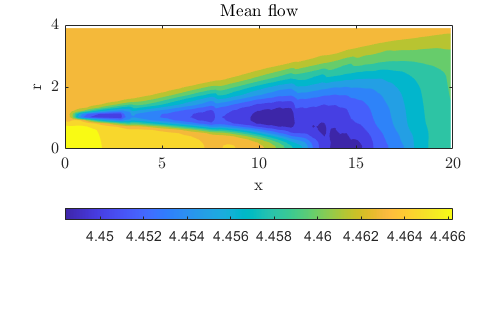}
	\includegraphics[width=6cm]{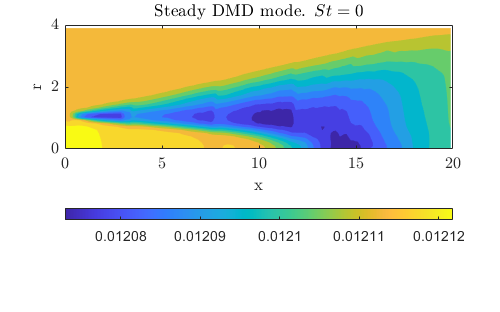}\\
    \includegraphics[width=6cm]{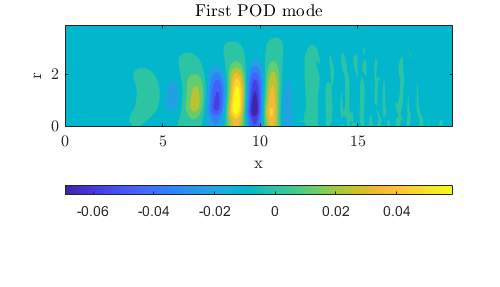}
    \includegraphics[width=6cm]{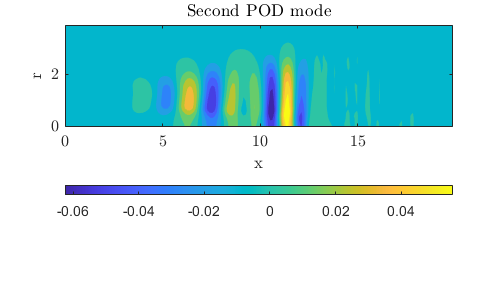}\\
    \includegraphics[width=6cm]{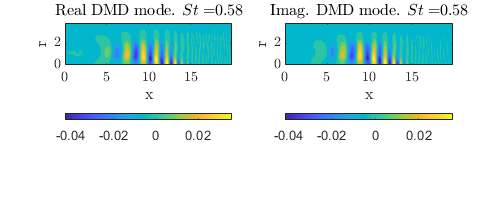}
	\includegraphics[width=6cm]{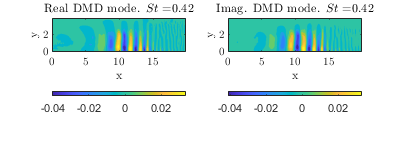}\\
	\caption{TC2: mean flow (top-left) and steady ($\St=0$) \DMD~mode (top-right).
	\POD~(middle) and \DMD~(bottom) dominant modes;
	from left to right: first and second highest energy/amplitude modes (\DMD~modal frequencies: St$=0.58,0.42$)}\label{fig:TC2_pod_dmd_modes}	
\end{figure}

Fig.~\ref{fig:TC2_pod_dmd_modes}~shows the mean flow and the most relevant \POD/\DMD~modes.
Again we observe the coincidence in shape of the mean flow and the \DMD~mode at $\St=0$.
Both the real \POD~modes and the complex \DMD~modes present features reminiscent of coherent wavepacket structures~\cite{towneEtAlSPOD}.

\begin{figure}[h]
	\centering
	\includegraphics[width=12cm]{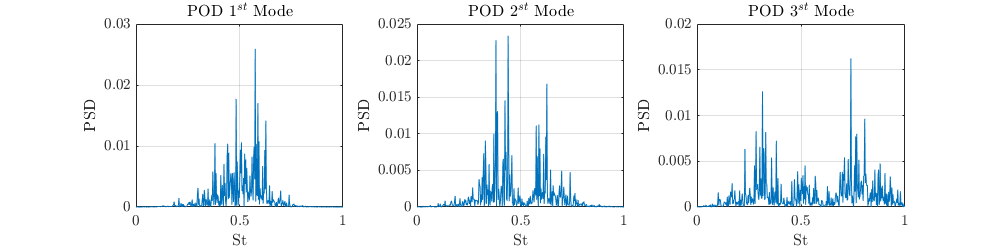}
	\caption{TC2: frequency contents corresponding to \POD~modes, obtained from the chronos matrix $\mathbf{R}_0$. 
	From left to right: first, second and third highest energy modes.}\label{fig:TC2_pod_spectrum}	
\end{figure}

\begin{figure}[H]
	\centering
	\includegraphics[width=8cm]{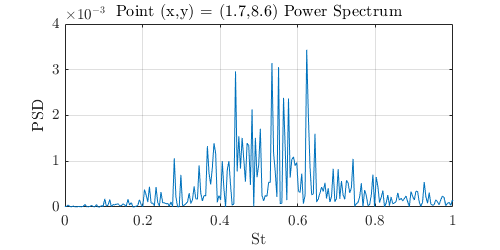}\\
	\includegraphics[width=8cm]{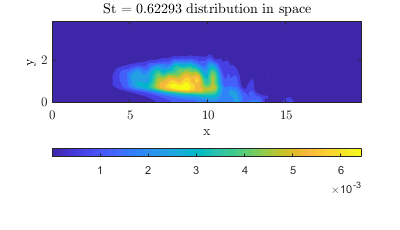}
	\caption{TC2: Frequency spectrum, computed using Eq.~\eqref{eq_psd} (top) and spatial distribution at an $St\approx0.6$ (bottom) obtained with \FFT.}\label{fig:TC2_fft}	
\end{figure}

Fig.~\ref{fig:TC2_pod_spectrum}	presents the frequency content of the first three dominant \POD~modes.
The spectral content of the \POD~modes is not as clean as those from TC1; 
rather, for each \POD~mode, all the frequencies are \textit{active}, which is to be expected in a complex, turbulent flow. It is worth mentioning that the dominant frequencies obtained through the spectrum in Fig. \ref{fig:TC2_pod_spectrum} coincide with the dominant frequencies of \DMD, Fig. \ref{fig:SV_dmdAmp_TC2}-right.
On the other hand, \DMD~is conceived so that \DMD~modes are associated with distinct, individual frequencies. 
However, that is of little help  in this case, recall the \textit{amplitudes} \Vs non-dimensional \textit{frequencies} distribution in Fig.~\ref{fig:SV_dmdAmp_TC2}.
The situation is not better for the \FFT~analysis, see Fig.~\ref{fig:TC2_fft}.	
In conclusion, as soon as datasets with certain complexity, data analysis tools with better performance become necessary.

\begin{figure}[H]
	\centering
	\includegraphics[width=6cm]{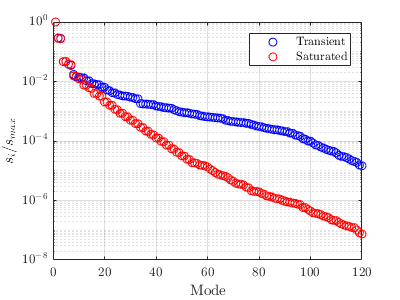}
	\includegraphics[width=6cm]{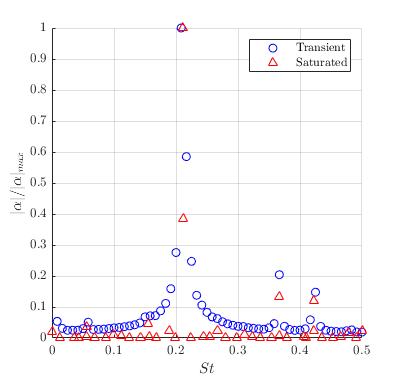}
	\caption{TC3 (both transient and saturated regimes). Left: singular values from the \POD~analysis. Right: \DMD~spectrum comparing non-dimensional frequencies with the \DMD~amplitudes, using with $\varepsilon=1\cdot 10^{-5}$ in Eq.~\eqref{eq_truncationStrategy}.}\label{fig:SV_dmdAmp_TC3}	
\end{figure}

\begin{figure}[H]
	\centering
	\includegraphics[width=6cm]{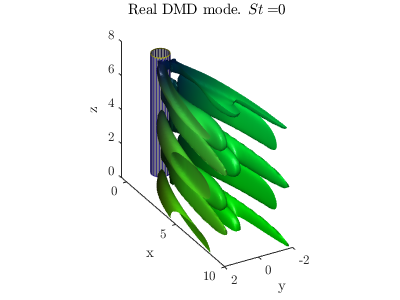}
	\includegraphics[width=6cm]{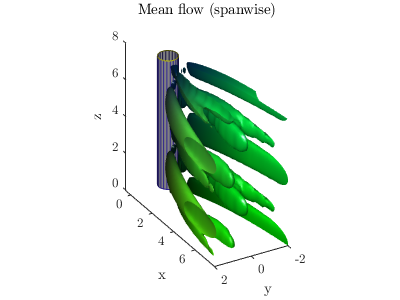}\\
    \includegraphics[width=6cm]{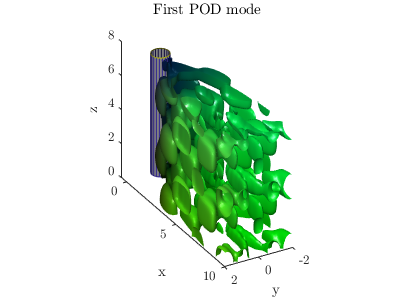}
    \includegraphics[width=6cm]{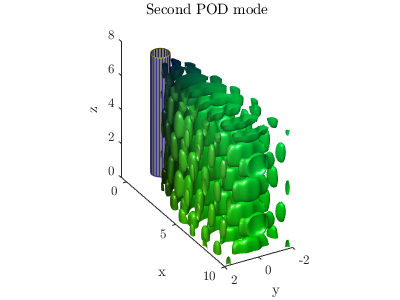}\\
    \includegraphics[width=6cm]{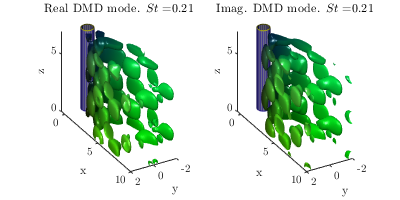}
	\includegraphics[width=6cm]{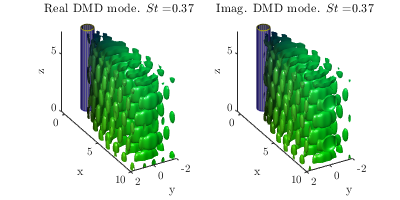}\\
	\caption{TC3 (saturated regime): mean flow (top-left) and steady ($\St=0$) \DMD~mode (top-right).
	\POD~(middle) and \DMD~(bottom) dominant modes;
	from left to right: first and second highest energy/amplitude modes (\DMD~modal frequencies: $St=0.21,0.37$). Iso-surfaces obtained for iso-value $9\cdot 10^{-4}$.}\label{fig:TC3_pod_dmd_modes}	
\end{figure}

As for TC3, Figure~\ref{fig:SV_dmdAmp_TC3}-left presents the singular values obtained by \POD~applied to the $z$-component of the velocity.
The singular value distributions are closer to that in TC1 than that in TC2;
in fact, singular values could be grouped in pairs again.
Note how, beyond the index $10$, the \textit{energy} of the \POD~modes is comparatively larger in the transient regime, as there are several flow structures that have not yet attenuated.
The \DMD~spectrum is also much simpler, and the modes at $\St\approx0.21$ and $0.36$ are easily identified.
This apparent simplicity is again a reflection of the laminarity of the dataset.
However, TC3 is more complex than TC1, and this is observed in the spatial support of the \POD/\DMD~modes shown in Fig.~\ref{fig:TC3_pod_dmd_modes}. 
The spanwise modulation of the modes is consistent with that of the flow solution, see Fig.~\ref{fig:cyl3D}.

\begin{figure}[h]
	\centering
	\includegraphics[width=12cm]{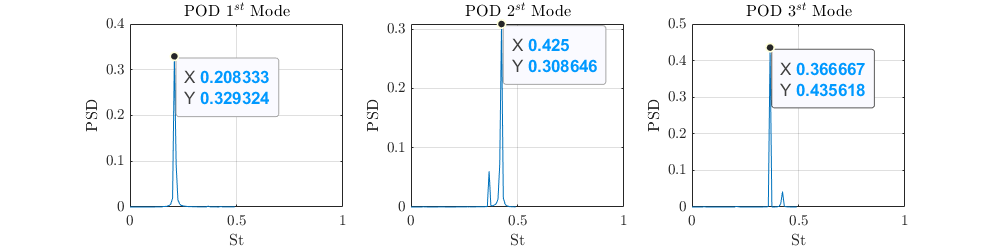}
	\caption{TC3: frequency contents corresponding to \POD~modes, obtained from the chronos matrix $\mathbf{R}_0$, Eq.~\eqref{eq_spatialReduction0}. 
	From left to right: first, second and third highest energy modes.}\label{fig:TC3_pod_spectrum}	
\end{figure}

Fig.~\ref{fig:TC3_pod_spectrum}	presents the frequency content of the first three dominant \POD~modes.
Again, the spectral content of the \POD~modes is relatively simple;
however, a single, distinct frequency cannot be assigned to each mode, as expected.
In fact, the second \POD~mode is linked to a frequency approximately double of $\St\approx0.21$, 
but a trace of $\St\approx0.36$ is also visible; 
the third \POD~mode is linked to $\St\approx0.36$ and yet is also contaminated by the first harmonic.

\begin{figure}[H]
	\centering
	\includegraphics[width=8cm]{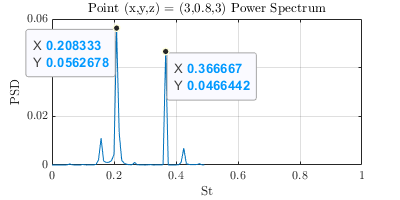}\\
	\includegraphics[width=8cm]{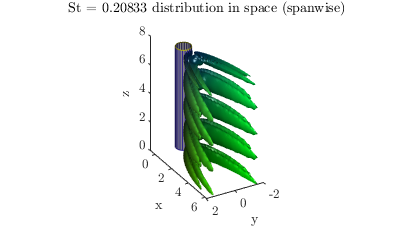}\\
	\caption{TC3 (saturated regime): Frequency spectrum, computed using Eq.~\eqref{eq_psd} (top);
	and spatial distribution at $\St\approx0.21$ 
	for iso-suface of spanwise velociy for $3\cdot 10^{-2}$ iso-value,
	obtained with \FFT.}\label{fig:TC3_fft}	
\end{figure}

The \FFT~analysis is summarized in Fig.~\ref{fig:TC3_fft}.
The spectrum at a tracer point reveals the components at $\St\approx0.21$ and $\St\approx0.36$, 
but contributions at other frequencies are also appreciable.
The spatial distribution of the horizontal \FFT~mode at $\St\approx0.21$ is different from that shown in Fig.~\ref{fig:FFT2DCyl}, \eg, it is slightly asymmetric. 
The spanwise modulation is specially visible in the \FFT~mode at $\St\approx0.21$ obtained from transversal velocity component of the dataset, Fig.~\ref{fig:TC3_fft}-bottom.



\subsection{Improving accuracy through redundancy: \SPOD~and \HODMD\label{sec:sPODandHODMD}}



We discuss now the results of the analysis using \SPOD~and \HODMD~methods to the datasets TC1-TC3.

\begin{figure}[H]
	\centering
    \includegraphics[width=9cm]{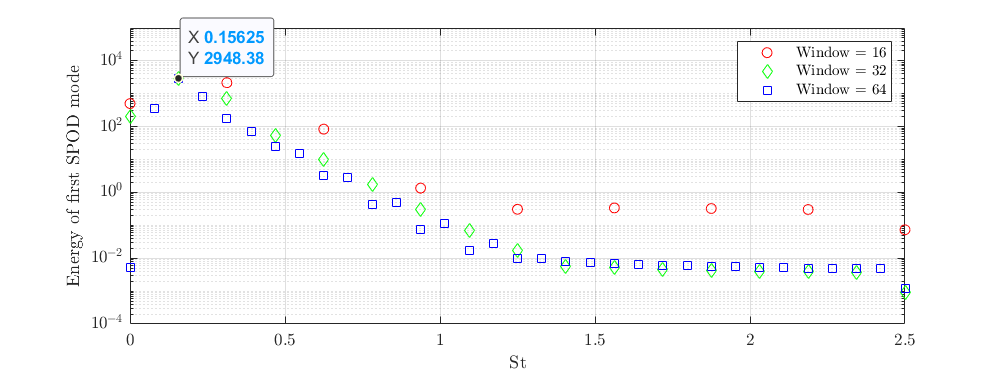}\\
    \includegraphics[width=4.5cm]{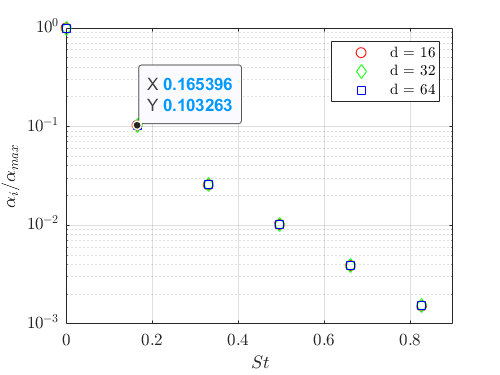}
	\includegraphics[width=4.5cm]{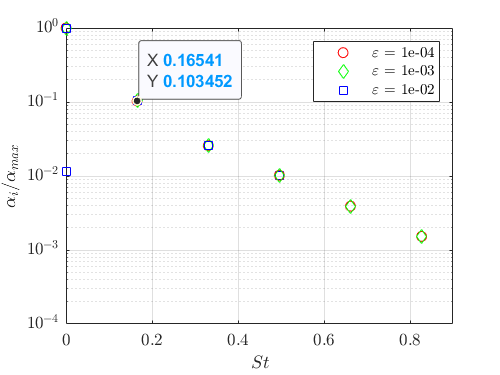}
	\caption{TC1, \SPOD~and \HODMD~analyses for window lengths that are powers of $2$. 
    Top, energy of the first \SPOD~mode \Vs~$\St$, and sensitivity to window length $n_t'$;
	bottom, \HODMD~$\alpha_i$ \Vs~$\St$ for different window lengths $d$ and tolerances $\varepsilon_{1}=\varepsilon_{2}$.}
	\label{fig:TC1_spod_hodmd_spectrum_spectralLeaking}
\end{figure}

\begin{figure}[H]
	\centering
    \includegraphics[width=9cm]{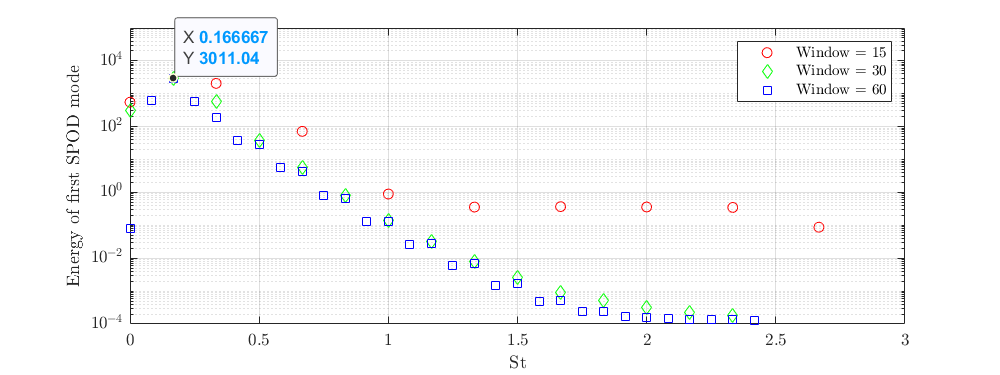}\\
    \includegraphics[width=4.5cm]{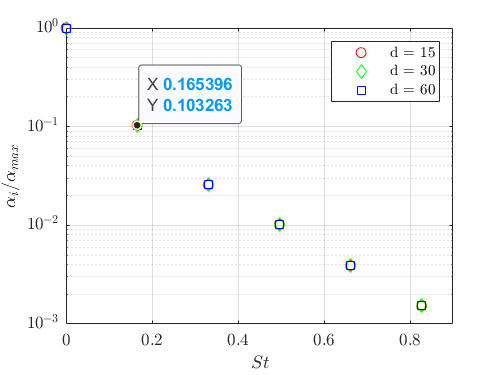}
	\includegraphics[width=4.5cm]{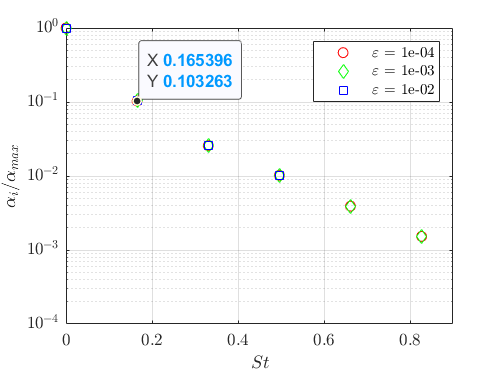}
	\caption{TC1, \SPOD~and \HODMD~analyses. Top, energy of the first \SPOD~mode \Vs~$\St$, and sensitivity to window length $n_t'$;
	bottom, \HODMD~$\alpha_i$ \Vs~$\St$ for different window lengths $d$ and tolerances $\varepsilon_{1}=\varepsilon_{2}$.}
	\label{fig:TC1_spod_hodmd_spectrum_notSpectralLeaking}
\end{figure}

Figures~\ref{fig:TC1_spod_hodmd_spectrum_spectralLeaking} and 
~\ref{fig:TC1_spod_hodmd_spectrum_notSpectralLeaking} present the \SPOD~(top) and \HODMD~(bottom) spectra. 
In particular, we investigate the influence of the window length (given by $n_t'$ for \SPOD~and $d$ for \HODMD) on the results obtained for TC1.
In a first comparison, we set window lengths that are powers of $2$, 
in order to obtain the optimal performance of the underlying \FFT~algorithm.  
Regarding the \SPOD~spectrum (Fig.~\ref{fig:TC1_spod_hodmd_spectrum_spectralLeaking}-top, 
obtained with an overlap of $50\%$ and the Hamming window function),
the first \SPOD~mode is most energetic at $\St\in\left(0.15,0.25\right)$, 
depending on the choice of the parameters.
This is a sign of spectral leaking~\cite{SPODGuideAIAAJ}, derived from the relatively short temporal sequence.
Regarding the \HODMD~results, Fig.~\ref{fig:TC1_spod_hodmd_spectrum_spectralLeaking} shows the sensitivity of the \HODMD~spectrum to window length (bottom left) and the tolerance in Eq.~\ref{eq_truncationStrategy}:
the higher the value, the less modes are retained in the reconstruction, and hence the fewer modes identified. 

\begin{figure}[H]
	\centering
	\includegraphics[width=6cm]{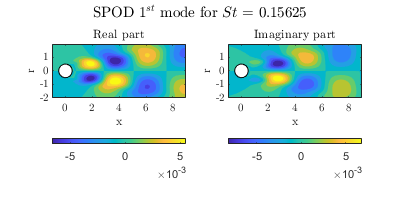}
	\includegraphics[width=6cm]{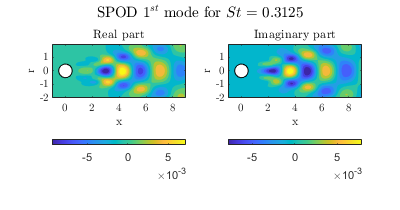}\\
    \includegraphics[width=6cm]{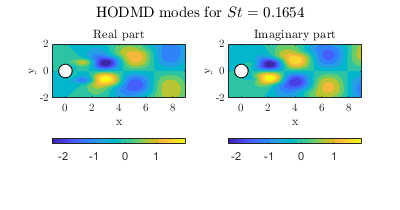}
	\includegraphics[width=6cm]{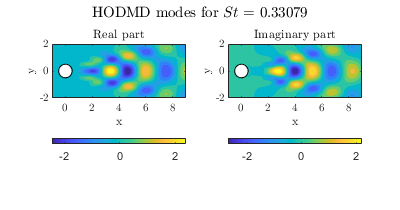}
	\caption{TC1, \SPOD~and \HODMD~most relevant (largest energy/amplitude, respectively) modes using $d = n_t'=64$.}\label{fig:TC1_hodmd_spod_modes_spectralLeakage}	
\end{figure}

\begin{figure}[H]
	\centering
	\includegraphics[width=6cm]{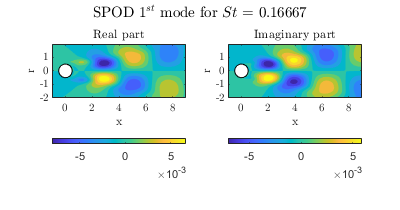}
	\includegraphics[width=6cm]{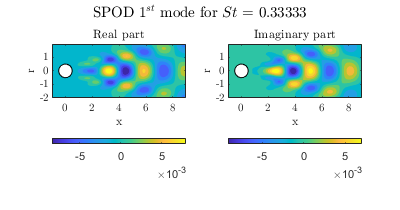}\\
    \includegraphics[width=6cm]{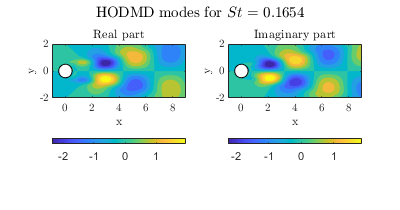}
	\includegraphics[width=6cm]{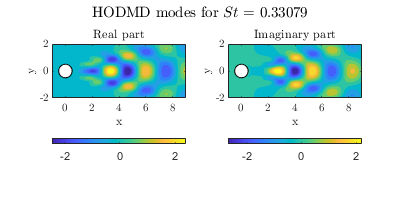}
	\caption{TC1, \SPOD~and \HODMD~most relevant (largest energy/amplitude, respectively) modes using $d = n_t'=60$.}\label{fig:TC1_hodmd_spod_modes_notSpectralLeakage}	
\end{figure}

\SPOD~retrieves the correct frequencies if window lengths multiple of $15$ are employed, as these leads to a set of bins coinciding with $\St\approx0.16$. 
These results are also summarized in Fig.~\ref{fig:TC1_spod_hodmd_spectrum_notSpectralLeaking}.
The \HODMD~results appear also in Fig.~\ref{fig:TC1_spod_hodmd_spectrum_notSpectralLeaking}.
The \SPOD~and \HODMD~modes 
are shown in Fig.~\ref{fig:TC1_hodmd_spod_modes_spectralLeakage} and Fig.~\ref{fig:TC1_hodmd_spod_modes_notSpectralLeakage}.
They look very similar to those obtained by \POD~and \DMD.

\begin{figure}[H]
	\centering
    \includegraphics[width=12cm]{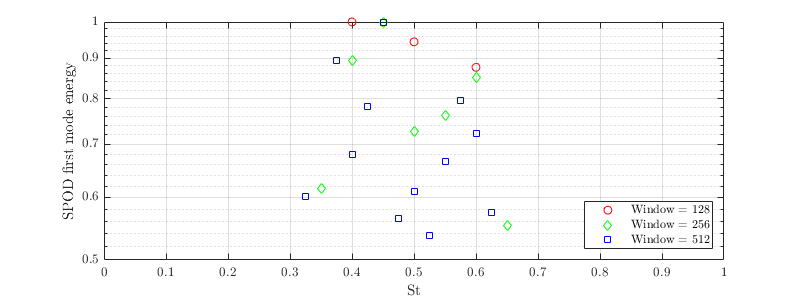}\\
    \includegraphics[width=6cm]{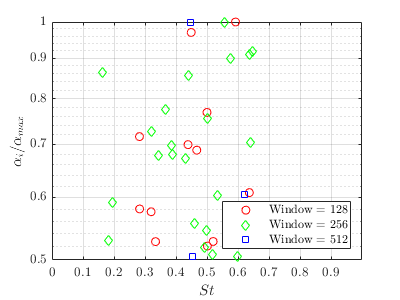}
	\includegraphics[width=6cm]{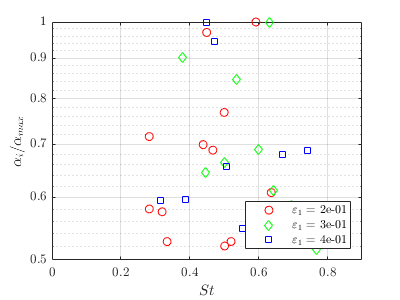}
	\caption{TC2, \SPOD~and \HODMD~analyses. Top, energy of the first \SPOD~mode \Vs~$\St$, and sensitivity to window length $n_t'$;
	bottom, \HODMD~$\alpha_i$ \Vs~$\St$ for different window lengths $d$ and tolerances $\varepsilon_{1}= 0.2$ and $ \varepsilon_{2}=0.5$.}
 
	\label{fig:TC2_spod_hodmd_spectrum}
\end{figure}

\begin{figure}[H]
	\centering
	\includegraphics[width=12cm]{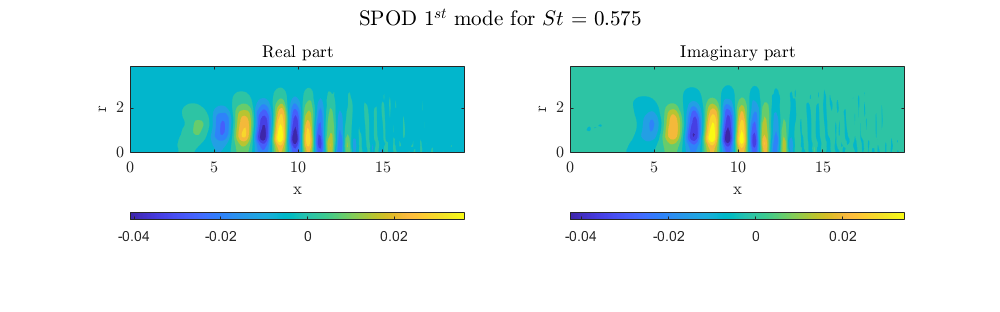}
	\includegraphics[width=12cm]{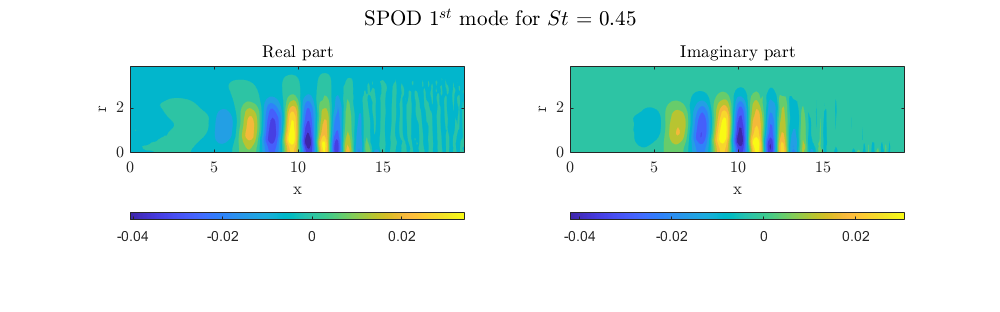}\\
    \includegraphics[width=12cm]{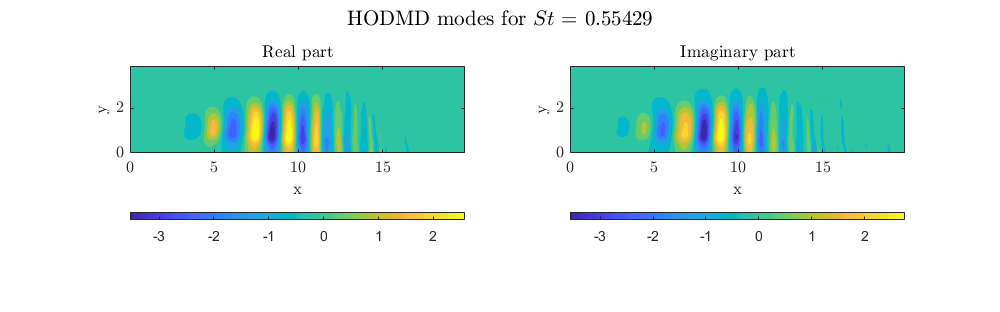}
	\includegraphics[width=12cm]{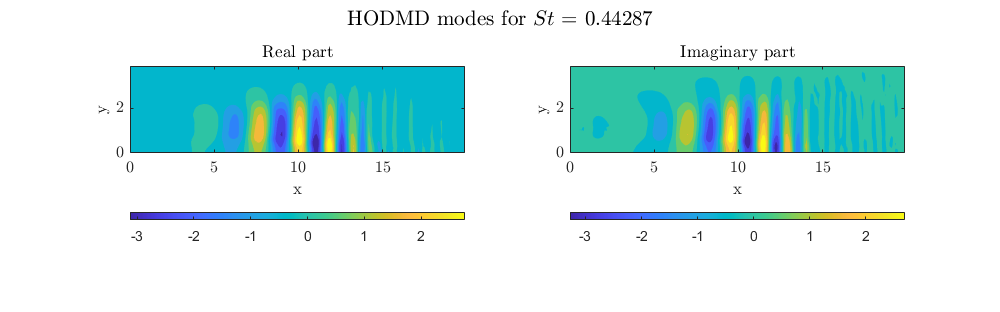}
	\caption{TC2, \SPOD~and \HODMD~most relevant (largest energy/amplitude, respectively) modes.}\label{fig:TC2_hodmd_spod_modes}	
\end{figure}

The spectra for the turbulent jet dataset (TC2) are summarized in Fig.~\ref{fig:TC2_spod_hodmd_spectrum}.	
This dataset is much longer ($n_t=5000$), which allows for more flexibility in the choice of the \SPOD~window length $n_t'$;
again we consider powers of $2$.
Fig.~\ref{fig:TC2_spod_hodmd_spectrum} (top) confirms that the most energetic \SPOD~mode is most active at $\St\approx0.6$.	
Fig.~\ref{fig:TC2_hodmd_spod_modes} (bottom) shows the largest amplitude \HODMD~modes and its sensitivity to $d$ and $\varepsilon_1=\varepsilon_2$.
In this case, the modes retrieved cover a range of $\alpha_i/\alpha_{max}\in\left(0.5, 1\right)$.
For values of $d=128-256$ and $\varepsilon_{1}=\varepsilon_{2}\in \left(0.2, 0.3\right)$, the most relevant mode appears at the expected $\St\approx0.6$.
The spatial support of both the \SPOD~and \HODMD~modes 
is shown in Fig.~\ref{fig:TC2_hodmd_spod_modes},
which again are very similar to those obtained by \POD~and \DMD.

\begin{figure}[H]
	\centering
    \includegraphics[width=12cm]{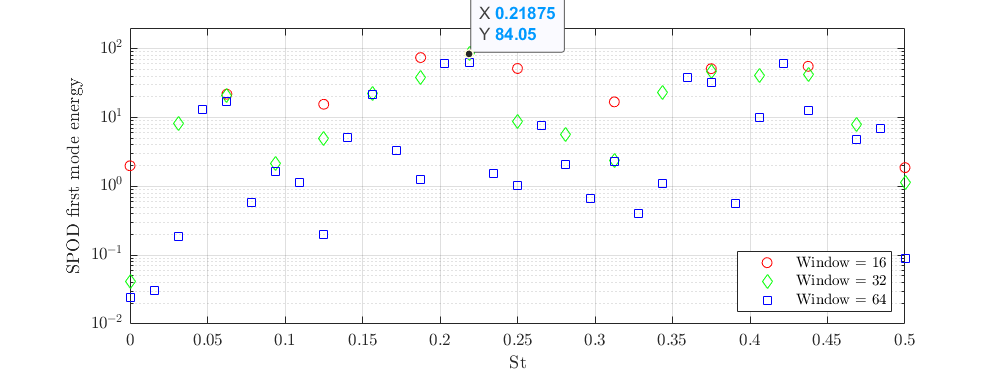}\\
    \includegraphics[width=6cm]{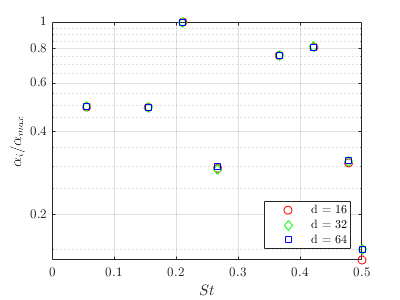}
	\includegraphics[width=6cm]{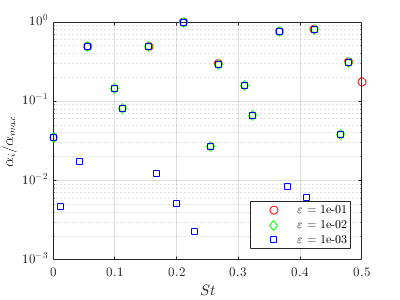}
	\caption{TC3 (saturated regime), \SPOD~and \HODMD~analyses. Top, energy of the first \SPOD~mode \Vs~$\St$, and sensitivity to window length $n_t'$;
	bottom, \HODMD~$\alpha_i$ \Vs~$\St$ for different window lengths $d$ and tolerances $\varepsilon_{1}=\varepsilon_{2} = 1\cdot 10^{-5}$.
 }
	\label{fig:TC3_spod_hodmd_spectrum_2}
\end{figure}

\begin{figure}[H]
	\centering
	\includegraphics[width=12cm]{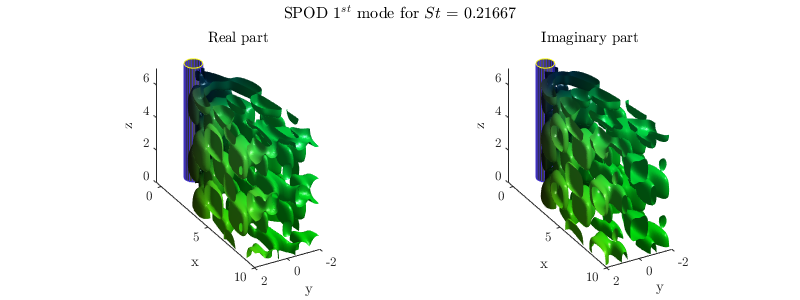}
	\includegraphics[width=12cm]{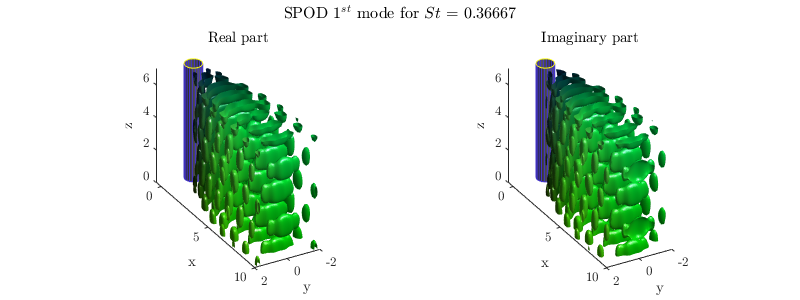}\\
    \includegraphics[width=12cm]{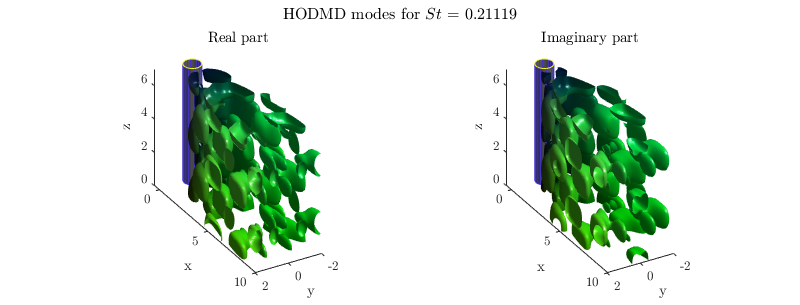}
	\includegraphics[width=12cm]{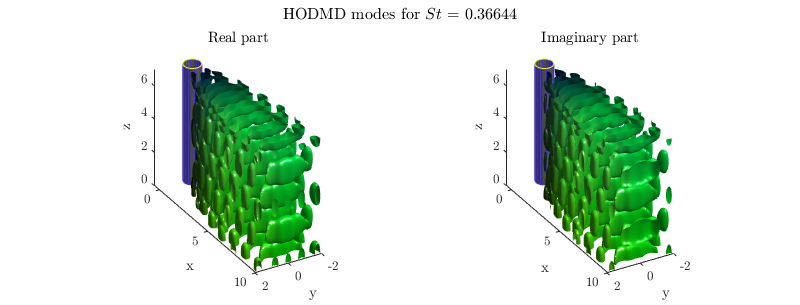}
	\caption{TC3 (saturated regime), \SPOD~and \HODMD~most relevant (largest energy/amplitude, respectively) modes. Iso-surface of spanwise velocity for $2.8\cdot 10^{-1}$ iso-values.}\label{fig:TC3_hodmd_spod_modes}	
\end{figure}

Regarding the three-dimensional laminar cylinder flow (TC3), 
the study is conducted on sections of the dataset of $n_t=120$, 
following \cite{bookLeClaincheHODMD}.
In this case, the target frequency is mainly $\St\approx0.21$ and $\St\approx0.36$,
and thus we will consider window lengths up to $64$ snapshots, 
so that the resulting bins can capture that frequency.
Fig.~\ref{fig:TC3_spod_hodmd_spectrum_2} (top) shows the most energetic \SPOD~mode. For a window length $n_t'=64$, 
this mode has largest energy at $\St\approx0.21$, 
also $\St\approx0.36$ seems to be relevant.
For window lengths $n_t'=16$ and $32$, the bin resolution is not enough for singling out the frequency of interest.
Fig.~\ref{fig:TC3_hodmd_spod_modes} (bottom) shows again the largest amplitude \HODMD~modes and its sensitivity to $d$ and $\varepsilon_1=\varepsilon_2$.
In this case, the modes retrieved cover a range of $\alpha_i/\alpha_{max}\in\left(10^{-3}, 1\right)$;
the most relevant modes appear at the frequencies $\St\approx0.21$ and $0.36$.
The spatial support of both the \SPOD~and \HODMD~modes 
is shown in Fig.~\ref{fig:TC3_hodmd_spod_modes},
which again are very similar to those obtained by \POD~and \DMD.\\
The advantages of using \SPOD~and \HODMD~in complex flows compared to classical \POD, \DMD~or \FFT~is clear. 
Both \SPOD~and \HODMD~offer cleaner and more accurate results that their standard counterparts. 
This is very useful to develop \ROMs, capable to reconstruct the original datasets and suitable for data forecasting (see Refs.~\cite{LeClaincheVegaPoF17, kouLeClaincheHODMDCriterion, le2018wind}).
Finally, as we investigated the performance of these two methods on both, saturated  (see Figures~
\ref{fig:TC3_spod_hodmd_spectrum_2} and~
\ref{fig:TC3_hodmd_spod_modes}) and transient regimes (results are not included for their similarity to the saturated case, as well as for brevity) of TC3. We can conclude that the advantage of \HODMD~is in the two tolerances of the algorithm, which are capable to reduce the data dimensionality, to remove spatial redundancies, and then to retain the most relevant spectral modes, as well as the luck of spectral leakage which is caused by \DFT~in \SPOD.
\newpage
\vskip-4.0cm

\clearpage
\newpage

\clearpage
\newpage

\clearpage
\newpage


%
%
%
%
%

\subsection{Feature detection based on multi-resolution analyses: mPOD and mrDMD\label{sec:mPOD}}


The multi-resolution techniques \mPOD~and \mrDMD, extensions of \POD~and \DMD~respectively, are suitable to identify different spatio-temporal scales defining the flow dynamics. Their application should be carried out in complex flows. 
The window size and the frequency filter selected in both methodologies is crucial to be able to progressively differentiate fast and low frequencies, properly modelling the flow physics. These methods, and specially \mrDMD, require a careful calibration, with the aim at obtaining robust results, and to be capable to differentiate the fast physical frequencies from unrealistic fast frequencies, which are erroneously captured by the method mixing chaotic dynamics with the periodic or quasi-periodic physical modes driving the flow. 

\begin{figure}[H]
    \subfloat[\label{fig:mPODCyl2D} ] {\includegraphics[width=1.0\textwidth]{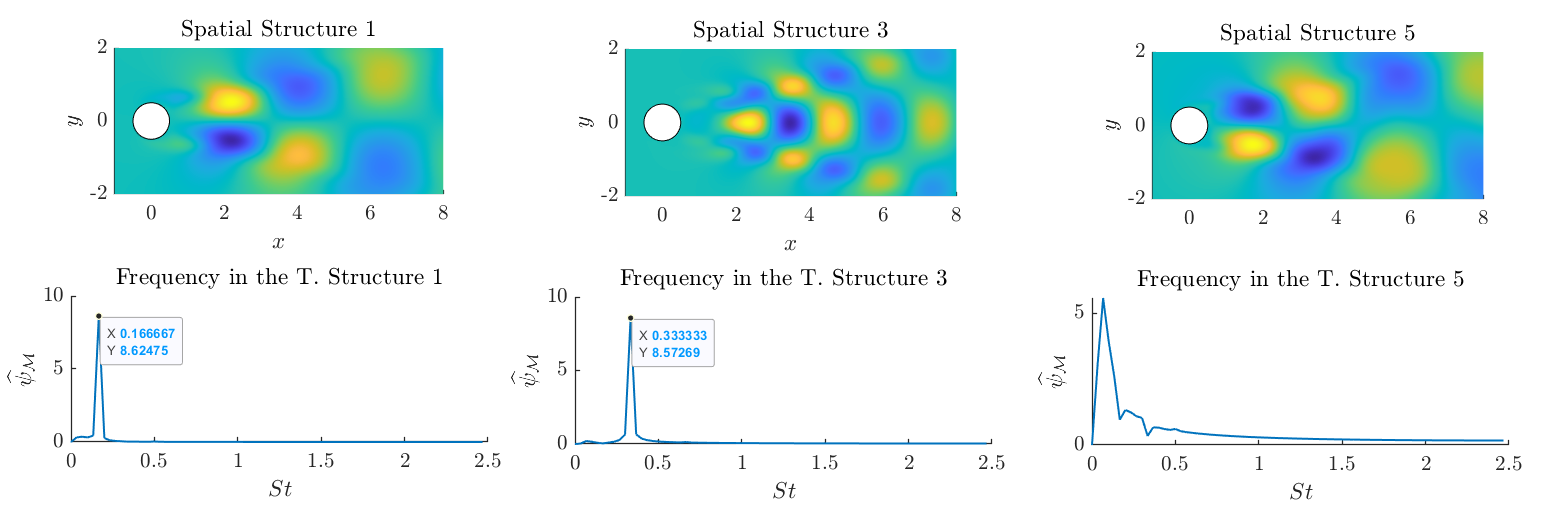}}\\
    \subfloat[\label{fig:mPODJet} ] {\includegraphics[width=1.0\textwidth]{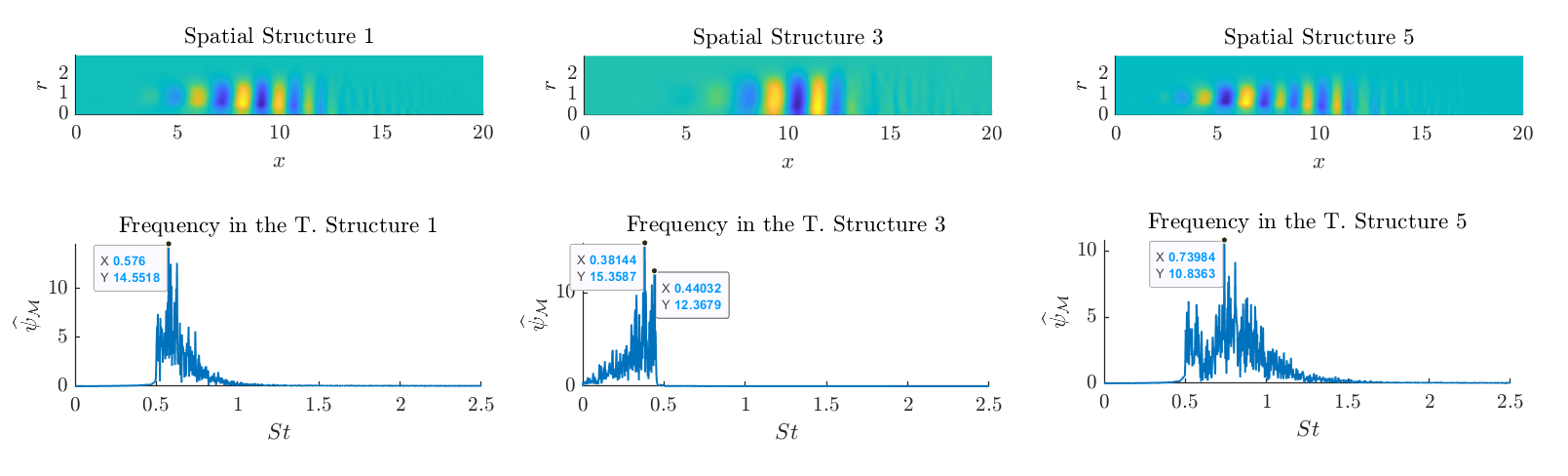}}\\
    
    \caption{ (a): TC1, \mPOD~modes (first row) and their corresponding frequency spectrum (second row). From left to right: from most to less energetic modes, $f_c=[0.2, 0.25]$. (b): TC2, \mPOD~modes (first row) and their corresponding frequency spectrum (second row), $f_c=[0.45, 0.5]$.
    }\label{fig:mPOD2DJet}
\end{figure}
\begin{figure}[H]
    
     \subfloat[\label{fig:mPODCyl3D_1} ] {\includegraphics[width=1.0\textwidth]{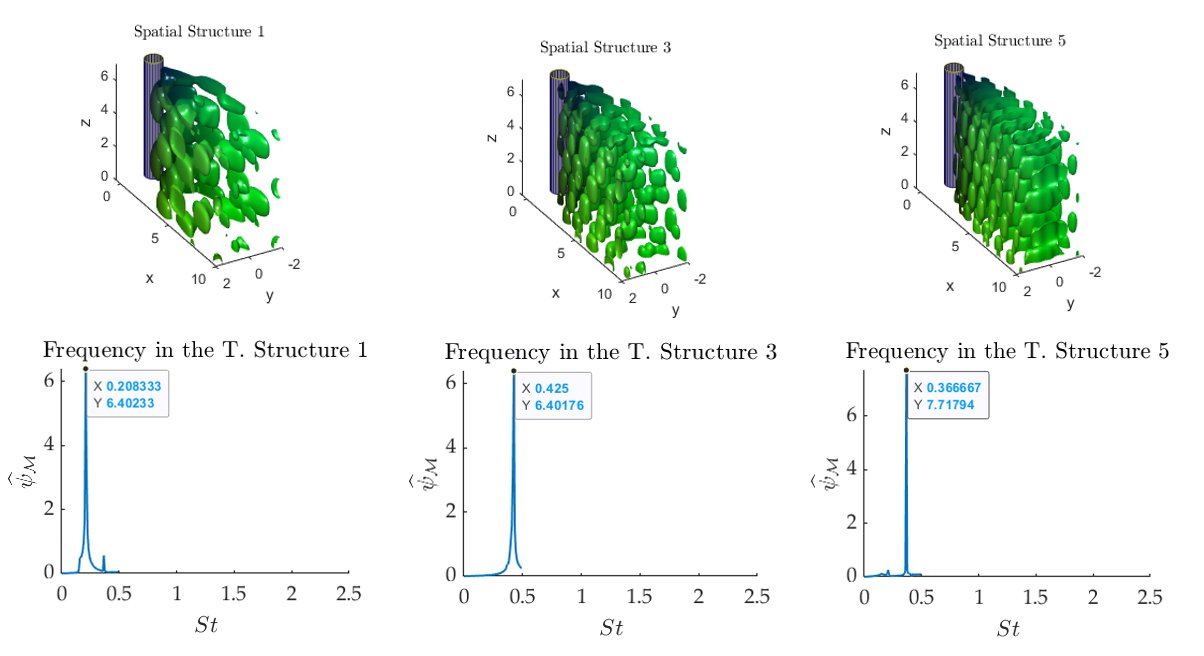}}\\
    
    \caption{ TC3: (saturated regime). \mPOD~modes (first row) and their corresponding frequency spectrum (second row). From left to right: from most to less energetic modes, $f_c=[0.15, 0.40]$
    . 
    }\label{fig:mPODCyl3D}
\end{figure}

Figures~\ref{fig:mPOD2DJet} and~\ref{fig:mPODCyl3D} show the three most energetic \mPOD~modes and their corresponding frequencies for all three testcases, as well as filter banks $\left (f_c\right )$. It is worth mentioning that in order to choose a suitable filter bank, prior knowledge about the flow is necessary.

%


However, in the \mPOD~case, to properly identify the main flow dynamics, and to differentiate spurious from physical modes, it is necessary to have a prior knowledge of the frequencies driving the flow. In this way, it is possible to establish a proper low-pass or high-pass filter frequency band, which is a part of the methodology.

Using multi-resolution analyses is recommendable when the user wants to identify specific frequencies from a highly complex spatio-temporal flow. These methods will be then able to differentiate fast and low events, representing physical solutions, or connected to specific noise or input (or external known) signals that for some specific reasons, it is worth to know with detail to characterize the flow.\\

\begin{figure}[H]
    \subfloat[\label{fig:TC1_mrDMD_spectrum} ] {\includegraphics[width=0.5\textwidth]{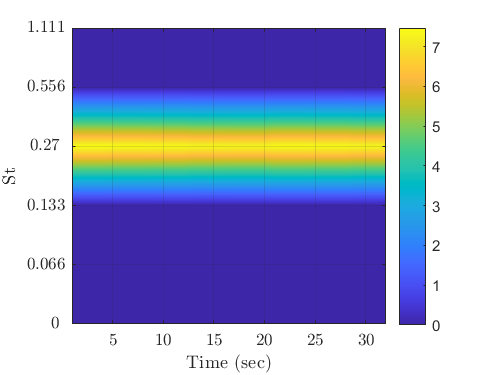}}
    \subfloat[\label{fig:TC2_mrDMD_spectrum} ] {\includegraphics[width=0.5\textwidth]{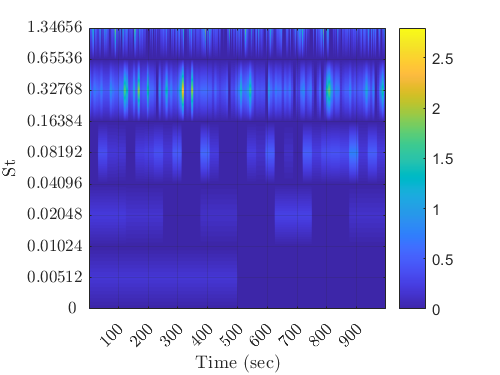}}\\
    
     \subfloat[\label{fig:TC3_mrDMD_spectrum_c} ] {\includegraphics[width=0.5\textwidth]{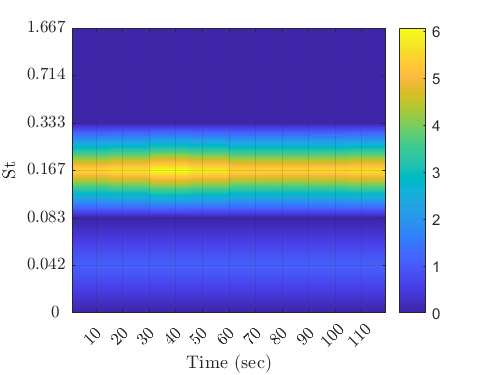}}
     \subfloat[\label{fig:TC3_mrDMD_spectrum_d} ] {\includegraphics[width=0.5\textwidth]{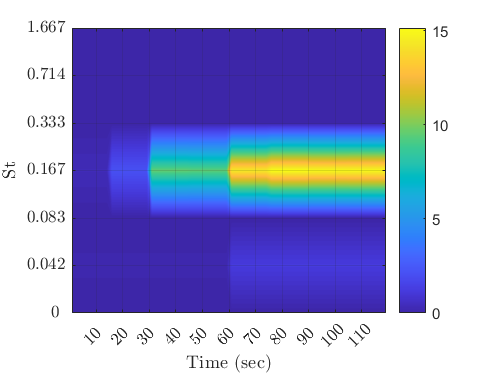}}\\
    

    \caption{\mrDMD~spectrum for TC1 (a), TC2 (b) (it is difficult to understand the dynamics in this testcase), TC3, saturated regime (c) and TC3, transient regime (d).}\label{fig:mrdmd_spectrum}
    
\end{figure}

The \mrDMD~spectra are shown in Fig. \ref{fig:mrdmd_spectrum}. For TC1 and TC3 it is easy to identify dominant frequency but for TC2 we have several ranges of dominant frequency. The advantage of \mrDMD~is that we have information not only in frequency but also in time. Although we need to have prior knowledge of flow dynamics in order to calibrate \mrDMD~hyper-parameters.


The spatial modes of all testcases are visually indistinguishable of those shown in Figures \ref{fig:PODmodesDMDmodes2DCyl}, \ref{fig:TC2_pod_dmd_modes}, \ref{fig:TC3_pod_dmd_modes},  and -for the sake of brevity- are not shown here.

\subsection{Beyond classical applications: resolvent analysis}\label{sec:RA}





In this section we explore the performance of the data-driven implementation of the Resolvent Analysis \RA~ on the three testcases investigated in this work.\\

\begin{figure}[H]
    \subfloat[\label{fig:raSpectrum2D}] {\includegraphics[width=6cm]{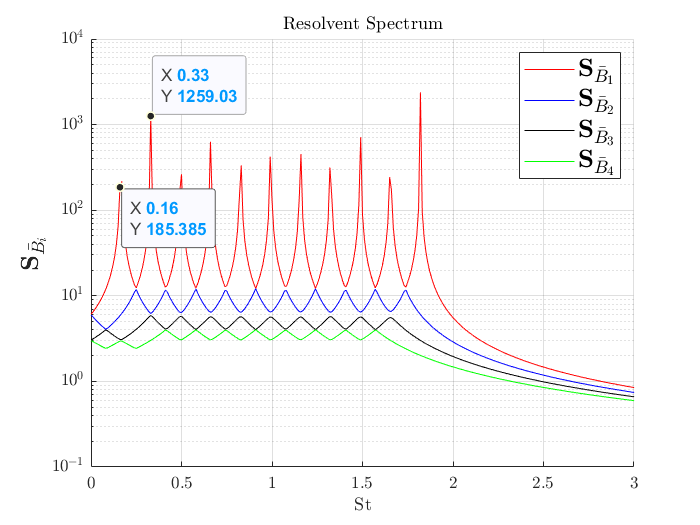}}\\
    
    \subfloat[\label{fig:raSpectrumJet_b} ] {\includegraphics[width=0.5\textwidth]{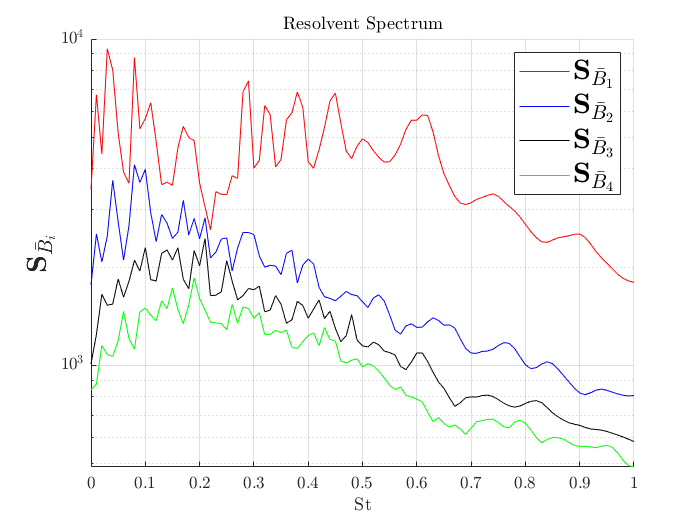}}
     \subfloat[\label{fig:raSpectrumJet_c} ] {\includegraphics[width=0.5\textwidth]{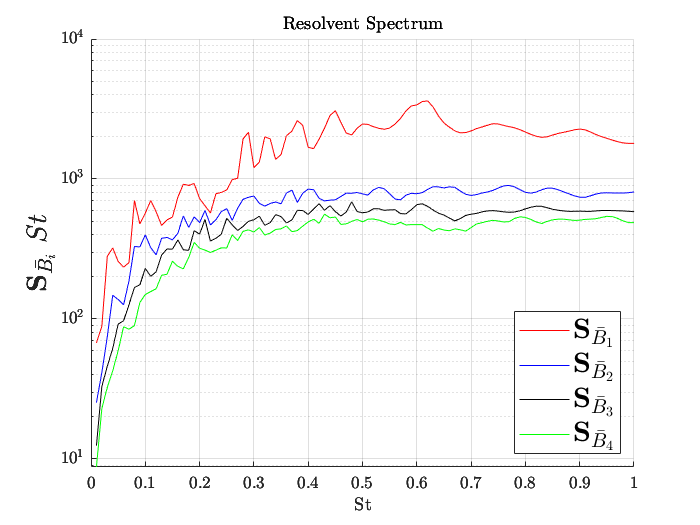}}\\
     
     \subfloat[\label{fig:raSpectrum3D_d} ] { \includegraphics[width=0.5\textwidth]{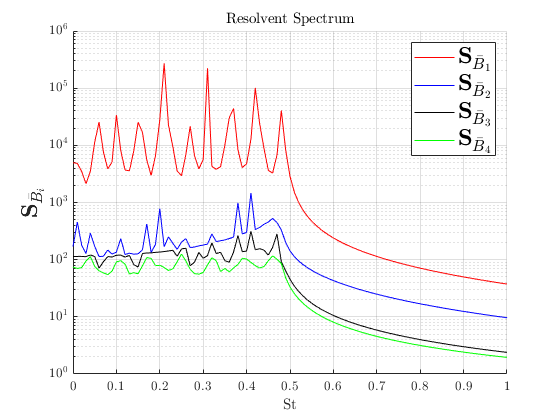}}
     \subfloat[\label{fig:raSpectrum3D_e} ] {\includegraphics[width=0.5\textwidth]{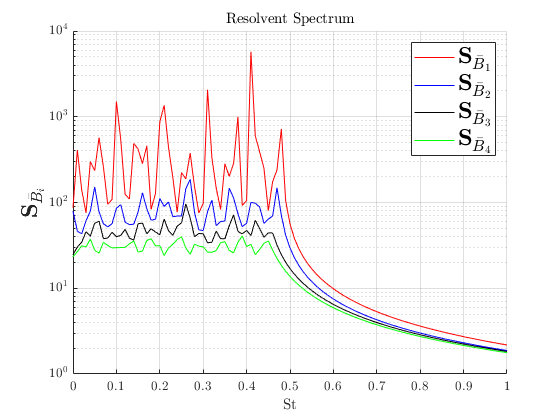}}\\
    
    \caption{ Resolvent analysis spectrum comparing the non-dimensional frequency St with the $\mathbf{S}_{\bar{B}}$ (see Eq.~\eqref{eq_resolvent_svd}) for (a) TC1, (b) and (c) TC2, (d) and (e) TC3 in transient and saturated regime, respectively.
    } \label{fig:raSpectrum}
    
\end{figure}

Figure. \ref{fig:raSpectrum} shows how the \RA~technique is capturing the frequencies in all testcases. Regarding TC1, and as can be seen in the Fig. \ref{fig:raSpectrum2D}, the \RA~was capable of capturing the wanted frequency (0.16), but not as the dominant frequency, as it identified two other harmonics as dominant frequencies. We can notice similar behavior in TC2 as well, where \RA~captures the relevant frequencies, but it also gave more importance to the smaller frequencies (see Fig. \ref{fig:raSpectrumJet_b}), and more accurate results can be obtained only when the frequencies are re-scaled (through the multiplication by $St$ ) as shown in Fig. \ref{fig:raSpectrumJet_c}. Same comments can also be applied to the results of TC3. Once again, the relevant frequencies are captured in both, the transient regime (Fig. \ref{fig:raSpectrum3D_d}) and saturated regime (Fig. \ref{fig:raSpectrum3D_e}), but other harmonics are also captured as dominant frequencies.\\

\begin{figure}[H]
    \subfloat[\label{fig:raModes2D}] {\includegraphics[width=\textwidth]{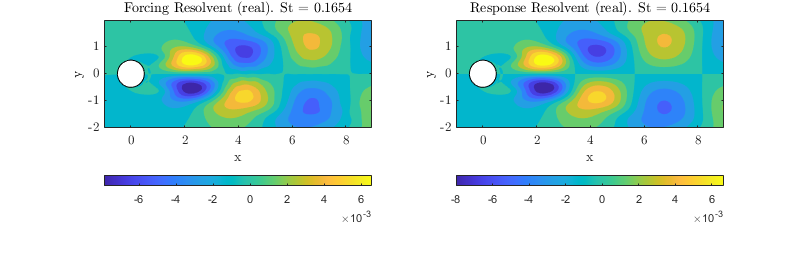}}\\
    
    \subfloat[\label{fig:raModesJet} ] {\includegraphics[width=\textwidth]{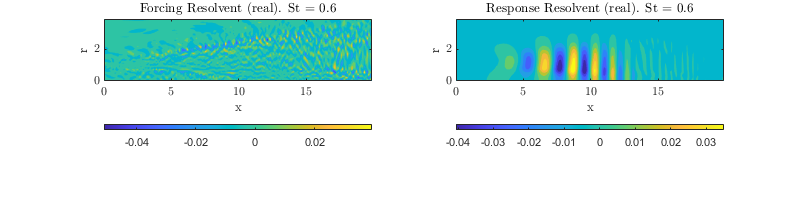}}\\
    
\subfloat[\label{fig:raModes_1_3D_b} ] {\includegraphics[width=0.5\textwidth]{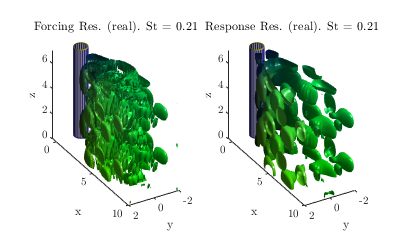}}
    \subfloat[\label{fig:raModes_2_3D_d} ] {\includegraphics[width=0.5\textwidth]{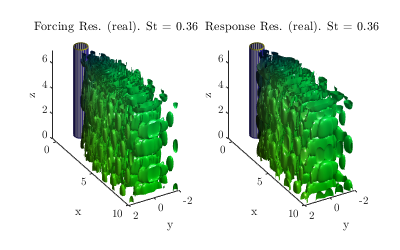}}\\
        
    \caption{Resolvent analysis modes for (a) TC1, (b) TC2, (c) and (d) TC3 in saturated regime, first and second Forcing and Response modes, respectively. 
    } \label{fig:RA_Modes}
    
\end{figure}

Figure. \ref{fig:RA_Modes} represents the modes obtained by the \RA~when applied to TC1, TC2 and TC3. In the case of TC1, as we can see in Fig. \ref{fig:raModes2D} the forcing and response modes are similar, and that is because the flow is saturated. Meanwhile in TC2, the growth rates are continuously changing, which will lead the difference between the forcing and response modes as seen in Fig. \ref{fig:raModesJet}, however the response mode is similar to the \DMD~and \POD~modes.\\

Regarding TC3, we have applied the \RA~on the saturated regime, where the growth rates are also changing, hence, the forcing and response modes are different as seen in Fig. \ref{fig:raModes_1_3D_b}, but again, the response mode is similar to the DMD mode. In this testcase in particular, we apply the \RA~algorithm to calculate the forcing needed to obtain the response of the \DMD~mode in a system.


%
%
%
%
%
%
%
%
%
%
%


\clearpage
\newpage

\section{Conclusions}\label{sec_conclusions}
In this work, we compare eight different modal decomposition techniques with the aim at introducing a critical assessment on what are the main benefits and constrains of each method as function of their application. This comparison involves three classical methods: Fast Fourier Transform (\FFT), Proper Orthogonal Decomposition (\POD) and Dynamic Mode Decomposition (\DMD), four extensions of the classical methods: Spectral \POD~(\SPOD), multi-scale \POD~(\mPOD), Higher Order \DMD~(\HODMD) and multi-resolution \DMD~(\mrDMD) and finally, the Resolvent Analysis (\RA). To the best of the authors' knowledge, this is the first time a review paper have gathered and compared all these modal decomposition techniques.\\

In order to provide an as fair as possible comparison, all the techniques have been applied to three testcases. The testcases  are coded as TC1, TC2 and TC3 and they are ordered with respect to their level of complexity. The first testcase considered was the velocity field of the laminar wake around cylinder at $Re_D = 100$, the second testcase is the pressure field of a turbulent jet flow at $Re_D = 10^6$ and the third and final testcase is the velocity field of the three dimensional transient wake around cylinder at $Re_D = 280$.
The first contribution in this work is a comparison between the performance of these methods in identifying the relevant frequencies when the databases are shortened. The obtained results imply the following: in cases where the data is limited, the \DMD-based methods outperform \FFT~and \POD-based methods. As \DMD-based methods deliver accurate results, mostly regardless of the size of the dataset and the window length, meanwhile \POD-based methods are tightened with some limitations as spectral leaking. Nevertheless, when taking larger amounts of data, all the methods deliver, in one way or another, comparable results. Furthermore, some of the methods fluctuate in frequencies based on the amount of data available, and this is linked to the flow complexity. As in these cases (TC2 and TC3), the methods are are exposed to failure in avoiding the modes connected to small flow scales.  \\
Next,the performance of these algorithms is thoroughly assessed concerning both the accuracy of the results and how these algorithms function with limited amounts of data. The obtained results are as follow: \\
As for the \textit{classical methods}, \POD, \DMD~and \FFT, all the methods were able to identify the dominant frequency as $\St\approx 0.16$ for TC1, but did not perform as good for TC2, as they all struggled in separating the relevant frequency at $\St\approx 0.6$ from other harmonics. For TC3, when \DMD~easily captured the main frequencies at $\St\approx 0.21$ and $\St\approx 0.36$, both \POD~and \FFT~were facing some difficulties in identifying the exact frequencies. Concerning the \textit{extension of the classical methods} We start with the results obtained from \SPOD~and \HODMD. The results obtained by the \HODMD~algorithm were quite accurate and robust, as it was able to capture the relevant frequencies in all testcases. For \SPOD, even though the techniques was able to capture the wanted frequencies, it was not an easy task. This technique, which usually deals with statistically stationary data, suffered from spectral leaking in all testcases. Regrading \mPOD~and \mrDMD, both methods identified the main frequency in TC1 without any trouble. However, a slight error in the frequencies was noticed in TC2 and TC3. As for the modes obtained by all the previously mentioned methods, they were all visually indistinguishable and consistent with the ones of the flow solution.

Finally, the \RA. When applying this technique to all testcases, the \RA~algorithm was able to capture the target frequencies, but not as dominant ones: there were always other modes that have been given more importance. Furthermore, the forcing and response modes are similar in TC1, because of the simplicity of the testcase, In the more complex testcases, the forcing and response mode are markedly different. Yet, the response modes are always similar to the modes obtained by the other techniques.\\
\vspace{6pt} 

\section*{Acknowledgment}
This work has been supported by SIMOPAIR (Project No. REF: RTI2018-097075-B-I00) funded by MCIN/AEI/10.13039/501100011033 and by the European
Union’s Horizon 2020 research and innovation program under the Marie Skłodowska-Curie Agreement number 101019137— FLOWCID. 
S.L.C. acknowledges the grant PID2020-114173RB-I00
funded by MCIN/AEI/10.13039/501100011033. 
The authors also acknowledge the support provided by Grant TED2021-129774B-C21, funded by MCIN/AEI/10.13039/501100011033 and by the European Union "NextGenerationEU"/PRTR.
\clearpage
\newpage
\appendix
\section{On the \DMD~amplitudes}\label{app_amplitudes}
\subsection{Computing the \DMD~amplitudes}

Since the introduction of the \DMD~method, a variety of strategies have been proposed 
to discriminate the relative importance of the dynamic modes identified. 
One of the earliest approaches, applied in the context of companion-matrix \DMD,  
simply compares the norm of the modes~\cite{rowleyEtAlJFM2009}. 

Other alternatives resort to the factorization model in Eq.~\eqref{eq_expansionDMD_matrixForm};
these approaches aim at identifying the \textit{amplitudes} $\alpha_i$.
One strategy obtains the $\alpha_i$ by projecting the first snapshot over the \DMD~basis~\cite{kutzEtAlBook} through Moore-Penrose pseudoinversion, $\boldsymbol{\alpha}=\boldsymbol{\Phi}^+\,\mathbf{v}_1$. 
Note that, since matrix $\mathbf{L}_0$ is unitary,
this step can be also accomplished at a reduced computational cost using the reduced chronos matrix:
\begin{eqnarray}\label{eq_alphaOptimizationSnapshot1}
    \boldsymbol{\alpha}=\boldsymbol{\Psi}^+\,\mathbf{c}_1.
\end{eqnarray}

\noindent Another alternative, rooted also on the factorization framework of Eq.~\eqref{eq_expansionDMD_matrixForm}, 
poses and solves a minimization problem in the Frobenius norm~\cite{jovanovic_SPDMD}: 
\begin{eqnarray}\label{eq_alphaOptimization}
  \min\limits_{\alpha_i}\| \mathbf{V}_{1}^{n_t-1} - \boldsymbol{\Phi}\mathbf{D}_{\alpha}\mathbf{V}_\mu\|^2_F,
\end{eqnarray}
\noindent Again, the fact that matrix $\mathbf{L}_0$ is unitary and does not affect the norm in Eq.~\ref{eq_alphaOptimization},
leads to the optimization problem:
\begin{eqnarray}\label{eq_alphaOptimization_projected}
  \min\limits_{\alpha_i}\| \mathcal{C}_1^{n_t-1} - \boldsymbol{\Psi}\mathbf{D}_{\alpha}\mathbf{V}_\mu\|^2_F.
\end{eqnarray}

\begin{figure}[H]
	\subfloat[\label{fig_amps_JovVsRowley}           ]{\includegraphics[width=7.5cm]{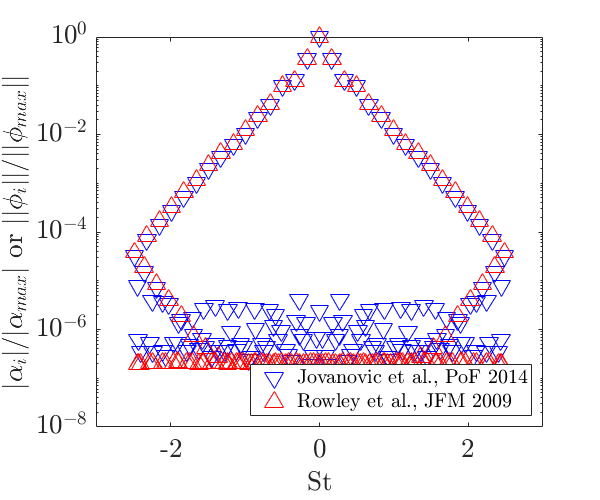}}
	\subfloat[\label{fig_amps_CompanionVsSimilarity} ]{\includegraphics[width=7.5cm]{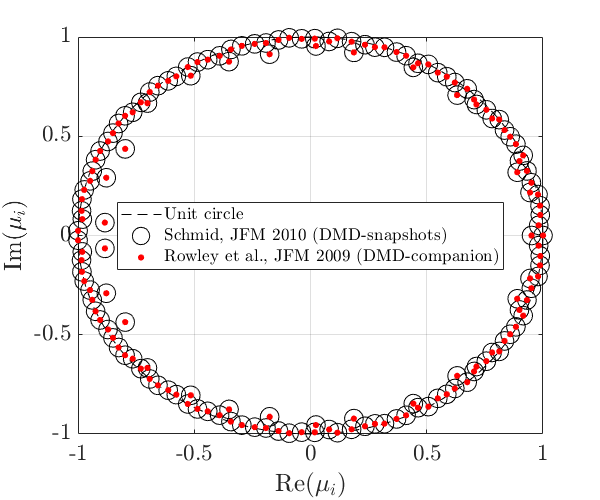}}\\
	\subfloat[\label{fig_amps_BKVsJov}               ]{\includegraphics[width=12cm]{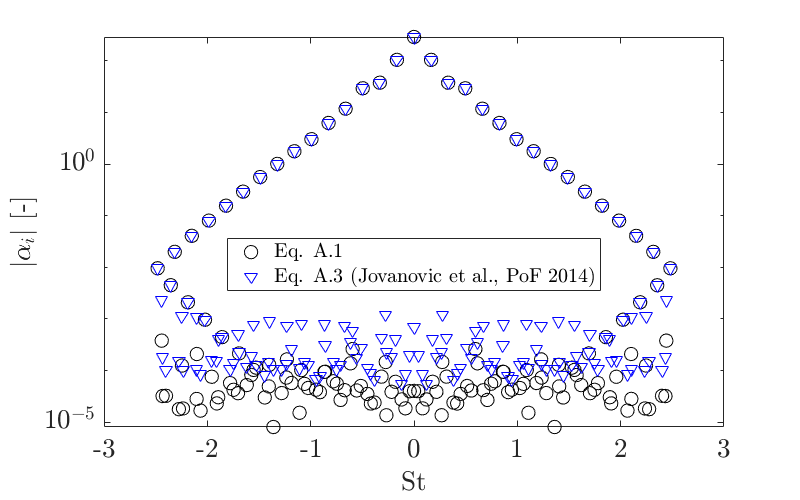}}
	\caption{Two-dimensional laminar wake around cylinder at $\Rey_D=100$: comparison of the relative importance of the different \DMD~modes, using different techniques.}\label{fig_amps}
\end{figure}

\noindent Figure~\ref{fig_amps} illustrates the concepts covered in the previous discussion 
using the TC1 dataset. 
Fig.~\ref{fig_amps_JovVsRowley} compares the relative importance of the \DMD~modes identified by 
both the companion and the similarity transformation based \DMD~algorithms,
using either the companion \DMD~mode norms~\cite{rowleyEtAlJFM2009} or the amplitudes obtained using~Eq.~\eqref{eq_alphaOptimization_projected}~\cite{jovanovic_SPDMD}.
In both cases, the frequency associated to the vortex shedding of the wake is clearly identified. 
Also, note how Fig.~\ref{fig_amps_CompanionVsSimilarity} confirms that both companion and the similarity transformation based \DMD~algorithms identify the same Ritz values.
Finally, Fig.~\ref{fig_amps_BKVsJov} illustrates the difference in amplitudes obtained by projecting simply on the first snapshot (Eq.~\eqref{eq_alphaOptimizationSnapshot1}) or on the whole data matrix (Eq.~\eqref{eq_alphaOptimization_projected}). 
Table~\ref{tab_amps_stats} offers complementary information on 
the reconstruction error $\min\limits_{\alpha_i}\| \mathcal{C}_1^{n_t-1} - \boldsymbol{\Psi}\mathbf{D}_{\alpha}\mathbf{V}_\mu\|^2_F$ achieved and 
the computational effort required for each approach, averaged for three different realizations.

\noindent In view of this analysis, in this contribution we have favoured the approach given in Eq.~\eqref{eq_alphaOptimization_projected} for the computation of the amplitudes.

\begin{table}[h]
\centering
\caption{\label{tab_amps_stats} Two-dimensional laminar wake around cylinder at $\Rey_D=100$: 
comparison of strategies to assess the relative importance of the different \DMD~modes,
using different techniques.}
\scalebox{0.6}{
		\begin{tabular}{c c c c}
		\toprule
			Method                                     & Reconstruction error  &     & Computing time $[s]$ \\
			\midrule                                                      
			Companion \DMD~\cite{rowleyEtAlJFM2009}    & N.A.                  &     & $4.72\times10^{-2}$      \\
			Eq.~\eqref{eq_alphaOptimizationSnapshot1}  & $1.64\times10^{-6}$   &     & $1.85$     \\
			Eq.~\eqref{eq_alphaOptimization_projected} & $5.85\times10^{-13}$  &     & $6.53\times10^{-4}$ \\
			\bottomrule
		\end{tabular}
		}
\end{table}

%

\subsection{Computing the \HODMD~amplitudes}
The amplitudes identification in the \HODMD~method requires solving 
an overdetemined system of equations given by (see~\cite{LeClaincheVegaSoria17,bookLeClaincheHODMD}):
\begin{eqnarray}\label{eq_hodmdAmplitudes}
    \mathbf{C}\,\boldsymbol{\alpha}=\mathbf{r},
\end{eqnarray}
where the coefficient matrix $\mathbf{C}$ is defined as:
\begin{eqnarray}
    \mathbf{C}=
    \begin{bmatrix}
    \boldsymbol{\Psi}'\\
    \boldsymbol{\Psi}'\,\mathbf{D}_\mu\\
    \vdots\\
    \boldsymbol{\Psi}'\,\mathbf{D}_\mu^{n_t-1}
    \end{bmatrix},
\end{eqnarray}
\noindent with $\mathbf{D}_\mu$ a diagonal matrix whose non-zero elements are the Ritz values $\mu_i$. 
The forcing term $\mathbf{r}$ is given by:
\begin{eqnarray}
    \mathbf{r}=
    \begin{bmatrix}
    \mathbf{c}_{1}\\
    \mathbf{c}_{2}\\
    \vdots\\
    \mathbf{c}_{n_t}\\
    \end{bmatrix}.
\end{eqnarray}

This linear system is solved in the least-squares sense, 
which de facto implies solving:
\begin{eqnarray}\label{eq_hodmdAmplitudesNew_1}
    \mathbf{C}^{H}\mathbf{C}\,\boldsymbol{\alpha}=\mathbf{C}^{H}\,\mathbf{r}.
\end{eqnarray}
This system is solved using pseudo-inversion  based on \SVD~decomposition.

For certain large databases, the memory footprint for computing the pseudo-inverse of  $\mathbf{C}$ can exceed the \texttt{RAM} available.
In those cases, the structure of $\mathbf{C}$ can still be exploited to obtain $\boldsymbol{\alpha}$.
Indeed, matrix $\mathbf{C}^H\mathbf{C}$ can be recast as:
\begin{eqnarray}\label{eq_hodmdAmplitudesNew_2}
    \mathbf{C}^{H}\mathbf{C}=\sum\limits_{k=1}^{n_t} (\mathbf{D}_\mu^{k-1})^{H}\boldsymbol{\Psi}'^H\, \boldsymbol{\Psi}'\,\mathbf{D}_\mu^{k-1},
\end{eqnarray}
whereas the product $\mathbf{C}^H\mathbf{r}$ turns out to be:
\begin{eqnarray}\label{eq_hodmdAmplitudesNew_3}
    \mathbf{C}^{H}\mathbf{r}=\sum\limits_{k=1}^{n_t} (\mathbf{D}_\mu^{k-1})^{H}\boldsymbol{\Psi}'^H\,\mathbf{c}_{k},
\end{eqnarray}

\begin{figure}[H]
    \centering
	\includegraphics[width=10cm]{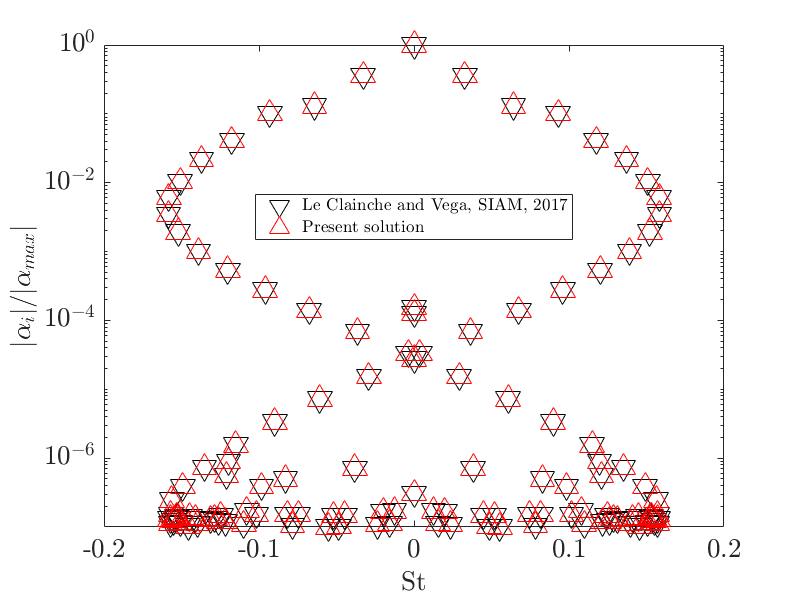}
	\caption{Two-dimensional laminar wake around cylinder at $\Rey_D=100$: comparison of the strategies to compute the \HODMD~amplitudes given in~\cite{leClaincheHODMDSIAM2017} and in Eqs.~\ref{eq_hodmdAmplitudesNew_1}-\ref{eq_hodmdAmplitudesNew_3} ($d=15$ and $\varepsilon_1=\varepsilon_2=10^{-14}$.}\label{fig_ampsHODMD}
\end{figure}

Fig.~\ref{fig_ampsHODMD} shows the result of this alternative strategy for TC1 case, showing that both strategies give the same result.

\subsection{Selecting the \DMD~amplitudes}

In section~\ref{sec_methodology} we discussed how, 
and differently from the eigenvalues obtained in \POD-based modes (\ie, $s_j=\sqrt{\lambda_j}$), 
the \DMD~amplitudes do not  have a direct interpretation as \textit{energies}.
This implies that \DMD~modes cannot be sorted rigourously according to moduli $\left|\alpha_i\right|$:
\eg, modes with large $\left|\alpha_i\right|$ might be decaying fast, 
\ie,  have large decrease rate $\Re{(\lambda_i)}\ll0$,
or alternatively, 
modes with moderate $\left|\alpha_i\right|$ might be long lived ($\Re{(\lambda_i)}\approx0$) 
or even unstable ($\Re{(\lambda_i)}>0$).

Several criteria have been derived to recognize such situations, see \eg~\cite{sayadiDMDSelection2015,kouDMDSelection2017}.
In this work, we consider the criterion proposed in~\cite{wavy_walls_garicano} to investigate transition in turbulent channels. 
The criterion sorts modes according to the following quantity:
\begin{equation}
    \gamma_i = |\alpha_i|\; \frac{e^{\Re (\lambda_i) \Delta t}-1}{\Re (\lambda_i)}.
\end{equation}

\begin{figure}[H]
	\centering
	\includegraphics[width=6cm]{./figures/results_3d/TC3_dmd_amplitudes_comparison}
	\includegraphics[width=6cm]{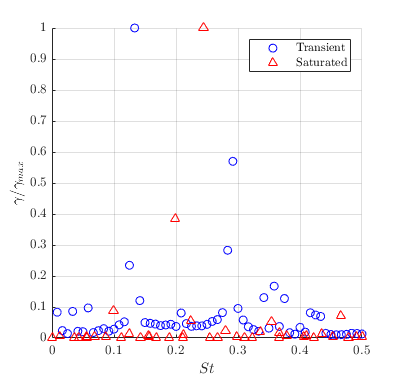}
	\caption{TC3, \DMD~analysis, $\varepsilon_{1}=1\cdot 10^{-5}$: $\alpha_i$ \Vs~$\St$ (left) and growth rates scaled \Vs~$\St$(right) for both the transient and saturated regimes.}\label{fig:DMD_amp_growth_comparison3D}	
\end{figure}

This criterion has been applied to the dataset TC3, for both the transient and saturated regimes.
Fig.~\ref{fig:DMD_amp_growth_comparison3D}

\begin{figure}[H]
	\centering
	\includegraphics[width=6cm]{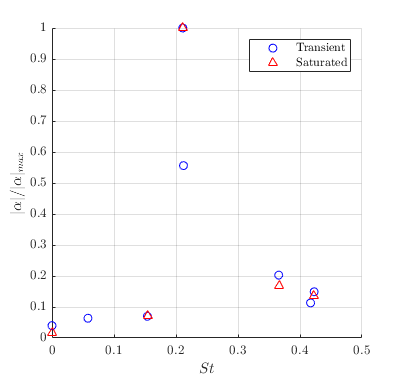}
	\includegraphics[width=6cm]{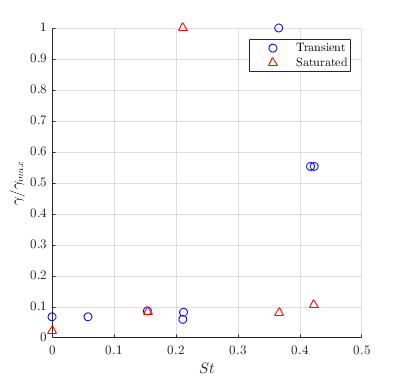}
	\caption{TC3, \DMD~analysis, $\varepsilon_{1}=5\cdot 10^{-2}$: $\alpha_i$ \Vs~$\St$ (left) and scaled growth rates \Vs~$\St$(right) for both the transient and saturated regimes.}\label{fig:DMD_amp_growth_comparison3D_2}	
\end{figure}

\clearpage
\newpage
\bibliography{Reference}

\end{document}

%% file: customizations.tex
\usepackage{epstopdf, epsfig}

\usepackage{orcidlink}

\usepackage{syntonly}

\usepackage{latexsym}                      
\usepackage[english]{babel}                
\usepackage{graphicx}
\usepackage{epstopdf}
\usepackage{subfig}
\usepackage{xfrac} 
\usepackage{hhline} 
\usepackage{xcolor} 

\usepackage{booktabs}                      
\usepackage{prettyref}                     
\usepackage{amssymb}                       
\usepackage{amsmath}                       
\usepackage{amsthm}                        
\usepackage{amstext}                       
\usepackage{array}                         
\usepackage{natbib} 
\usepackage{ifthen}                        
\usepackage{ifpdf}                         
\usepackage[]{color}               

\usepackage{appendix}

\usepackage{indentfirst} 
\usepackage{framed}                      %

\renewcommand{\Re}{\mathrm{Re}}
\renewcommand{\Im}{\mathrm{Im}}

\newcommand{\St}{\mathrm{St}}
\newcommand{\Rey}{\mathrm{Re}}
\newcommand{\M}{\mathrm{M}}

\newcommand{\DNS}{\textbf{DNS}}

\newcommand{\QR}{\textbf{QR}}
\newcommand{\POD}{\textbf{POD}}
\newcommand{\SVD}{\textbf{SVD}}

\newcommand{\TS}{\textbf{TS}}

\newcommand{\FFT}{\textbf{FFT}}
\newcommand{\DFT}{\textbf{DFT}}
\newcommand{\CSD}{\textbf{CSD}}
\newcommand{\DMD}{\textbf{DMD}}

\newcommand{\ROMs}{\textbf{ROM}s}

\newcommand{\HODMD}{\textbf{HODMD}}
\newcommand{\mrDMD}{\textbf{mrDMD}}
\newcommand{\RA}{\textbf{RA}}
\newcommand{\SPOD}{\textbf{SPOD}}
\newcommand{\mPOD}{\textbf{mPOD}}
\newcommand{\PSD}{\textbf{PSD}}

\newcommand{\MRA}{\textbf{MRA}}

\newcommand{\iu}{{i\mkern1mu}}

\usepackage[english]{babel}
\usepackage{tabularx}
\usepackage{tabulary}

\usepackage{footmisc}

\newcommand{\divg}[1]{\bnabla\cdot}

\newcommand{\Vs}{\textit{vs}~}
\newcommand{\etal}{\textit{et al.}}
\newcommand{\eg}{\textit{e.g.}} 
\newcommand{\ie}{\textit{i.e.}} 
\newcommand{\cfr}{\textit{cf.}} 

\usepackage{soul,xcolor}

\usepackage[many]{tcolorbox}
\newtcolorbox{cross}{blank,breakable,parbox=false,
              overlay={\draw[blue,line width=5pt] (interior.south west)--(interior.north east);
              \draw[blue,line width=5pt] (interior.north west)--(interior.south east);}}